\documentclass[11pt]{article}
\usepackage{amsmath}
\newcommand{\Qd}{Q^{\dagger}}

\newcommand{\be}{\begin{equation}}
\newcommand{\ee}{\end{equation}}
\newcommand{\bea}{\begin{eqnarray}}
\newcommand{\eea}{\end{eqnarray}}

\newcommand{\e}{{\rm e}}
\renewcommand{\d}{{\rm d}}
\begin{document} 
 
\title{Lattice fermion models with supersymmetry} 

\author{Paul Fendley$^1$, Bernard Nienhuis$^2$, and Kareljan Schoutens$^2$
\medskip \\ 
$^1$ Department of Physics, University of Virginia, \\
Charlottesville, VA 22904-4714 USA \\  
{\tt fendley@virginia.edu}\smallskip \\  
$^2$ Institute for Theoretical Physics,  
University of Amsterdam,\\
 Valckenierstraat 65, 1018 XE Amsterdam, The Netherlands \\  
{\tt nienhuis@science.uva.nl, kjs@science.uva.nl}}
\smallskip 
 
\maketitle 
 
\begin{abstract} 
\noindent
We investigate a family of lattice models with manifest ${\cal N}$=2
supersymmetry. The models describe fermions on a 1D lattice, subject
to the constraint that no more than $k$ consecutive lattice sites may
be occupied. We discuss the special properties arising from the
supersymmetry, and present Bethe ansatz solutions of the simplest
models. We display the connections of the $k=1$ model with the 
spin-$\frac{1}{2}$
antiferromagnetic XXZ chain at $\Delta=-1/2$, and the $k=2$ model
with both the $su(2|1)$-symmetric $tJ$ model in the ferromagnetic regime
and the integrable spin-1 XXZ chain at $\Delta=-1/\sqrt{2}$. We
argue that these models include critical points described
by the superconformal minimal models.

\end{abstract} 


 
\vskip .5in 
\section{Introduction: supersymmetric lattice models}

In the analysis of quantum mechanics and quantum field 
theory, one is often guided by the presence of symmetries. 
Among these symmetries, supersymmetry stands out as 
particularly powerful: supersymmetric theories enjoy special `magical' 
properties and in many cases supersymmetry leads to considerable 
computational simplification.

Supersymmetric field theories in two (2+0 or 1+1) dimensions are
particularly interesting. They have manifold applications, ranging from
their use as world sheet theories in (super-)string theory to their
role as continuum limits of critical models of 2D statistical
mechanics or 1D quantum chains.  For example, the field theory
describing the (experimentally realizable \cite{xxzexp})
antiferromagnetic XXZ spin chain at anisotropy $\Delta=-1/2$
has ${\cal N}$=(2,2) supersymmetry.

Other lattice
models, such as the two-dimensional tricritical Ising model along the
first-order transition line, are described by supersymmetric field
theories in the continuum limit \cite{SW} as well. 
In none of these cases is it obvious how the supersymmetry
manifests itself in the underlying lattice model. To
understand this issue, and to show how
supersymmetry improves our understanding of lattice models
directly, we introduced lattice models with explicit
supersymmetry \cite{FSd}. These models describe hard-core fermions on
a general lattice or graph, and the supersymmetry generators are
related to operators that create or annihilate these fermions. We
showed that the simplest of these models defined on a 1D chain is
critical, and is closely related to the XXZ chain at $\Delta=-1/2$.

In the present work, we greatly expand upon the work of \cite{FSd}. We
introduce a series of supersymmetric lattice models, the simplest one
that of \cite{FSd}.  Here we we limit our study to one-dimensional
(open or closed) chains, but the models (if not the results) can be
easily generalized to higher dimensions.  Analyzing the models in
detail, we pay particular attention to the following aspects: (i)
special features, such as zero-energy ground states, implied by the
supersymmetry, (ii) integrability by Bethe ansatz, (iii) mappings to
other models, and (iv) continuum limit and relation with models of
${\cal N}$=(2,2) superconformal field theory.  In section 2, we define
a series of models (denoted $M_k$), and compute their Witten index.
In section 3, we discuss the simplest model $M_1$ in detail, extending
the results of \cite{FSd}. In section 4, we show that the model $M_2$
at a particular coupling is closely related to the
$su(2|1)$-supersymmetric $tJ$ model, thus providing an intriguing
relation between spacetime supersymmetry and global symmetries with
fermionic generators (which are often called supersymmetries in the
condensed-matter literature).  We also show that at a different
coupling, $M_2$ is related to the spin-1 XXZ model with couplings
tuned to be integrable. The latter model and the model $M_1$ are
described in the continuum limit by the first two minimal models of
superconformal field theory. In section 5, we argue that a particular
coupling of the model $M_k$ is described by the $k$th superconformal
minimal model.

\section{A family of ${\cal N}$=2 supersymmetric lattice models}

\subsection{The models $M_k$: general structure}

To define the models $M_k$, we consider fermions on a 1D chain, 
subject to the condition 
that at the most $k$ consecutive lattice sites may be occupied.
This defines a Hilbert space ${\cal H}_k$, with subspaces 
${\cal H}^{(f)}_k$ defined by the condition that the eigenvalue
of the fermion-number operator $F$ be equal to $f$. The supersymmetry 
generators $Q=Q^-$ and $\Qd=Q^+$ act on the constrained 
Hilbert spaces ${\cal H}_k$ as
\be
Q^-: \ {\cal H}_k^{(f)} \rightarrow {\cal H}_k^{(f-1)} \ ,
\qquad
Q^+: \ {\cal H}_k^{(f)} \rightarrow {\cal H}_k^{(f+1)} 
\ee
so that 
\be
[F,Q^\pm]=\pm \, Q^\pm.
\label{eq:fq}
\ee
They are nilpotent:
\be
(Q^+)^2 = (Q^-)^2 = 0 \ .
\label{eq:qsquared}
\ee
One of the most important characteristics of models with spacetime
supersymmetry is that the algebra of the supersymmetry
generators involves the Hamiltonian $H$. For ${\cal N}$=2 supersymmetry,
the relation is
\be
H = \{ Q^+ , Q^- \} \ .
\label{eq:hamiltonian}
\ee
This definition, together with the nilpotency of the supersymmetry 
generators, implies that $H$ commutes with both the fermion
number and the supersymmetry generators
\be
[ F , H ] = 0 , \quad 
[ Q^+ , H ] =  [ Q^- , H ] = 0 \ .  
\label{eq:comH}
\ee
The algebraic structure summarized in the equations (\ref{eq:fq}),
(\ref{eq:qsquared}), (\ref{eq:hamiltonian}) and (\ref{eq:comH}) 
is characteristic of supersymmetric quantum mechanics
or ${\cal N}$=2 supersymmetric quantum field theory. 

To further specify the models $M_k$, we start by introducing 
creation and annihilation operators $c_i^\dagger$, $c_i$ for 
unrestricted fermions. They obey the usual anticommutation 
relations
$$
\{c_i,c_j^\dagger\}=\delta_{ij}\qquad\qquad
\{c_i,c_j\}=\{c_i^\dagger,c_j^\dagger\}=0,
$$ 
so there is at most one fermion per site.
The supersymmetry generators
are particular combinations 
of the fermion annihilation and creation operators. 
There is some freedom in
defining the $Q^\pm$ on ${\cal H}_k$ while maintaining the 
above algebraic structure. 
This freedom will lead to adjustable parameters $x_1,\ldots, x_{k-1}$
in the definitions of the models $M_k$.

For $k=1$, adjacent fermions are forbidden; i.e.\
the Hilbert space is constrained so that
nearest-neighbor sites on the lattice cannot be occupied. With this, 
the fermions can be thought of as hard-core dimers centered on 
site $i$, with the dimer covering the sites $i - \frac{1}{2}$ 
and $i + \frac{1}{2}$ on the dual lattice. 
Introducing the projection operator 
${\cal P}_i\equiv 1 - c^\dagger_i c_i$, we can write
$d_i \equiv {\cal P}_{i-1} c_i^\dagger {\cal P}_{i+1}$
and define the supersymmetry generators by
$$
Q^+ = \sum_{i=1}^{N} d^\dagger_i \ , 
\qquad 
Q^- = \sum_{i=1}^N d_i \ .
\label{eq:Qpm1}
$$
The resulting Hamiltonian will be displayed, and its eigenstates
found, in section \ref{sec:M1}.

Putting $k=2$, we can define
$d^\dagger_i \equiv {\cal P}_{i-1} c_i^\dagger {\cal P}_{i+1}$,
$e^\dagger_i \equiv 
{\cal P}_{i-2} c^\dagger_{i-1} c_{i-1} c_i^\dagger {\cal P}_{i+1}
+ {\cal P}_{i-1} c_i^\dagger c^\dagger_{i+1} c_{i+1} {\cal P}_{i+2}$
and define supersymmetry as
$$
Q^+=\sum_{i=1}^N 
[y_1 d^\dagger_i + y_2 e^\dagger_i] \ ,
\qquad
Q^-= (Q^+)^\dagger \ .
$$
The condition $(Q^\pm)^2=0$ is satisfied for all $y_1$, $y_2$,
leaving the ratio $x_1=y_1/y_2$ as a non-trivial adjustable parameter.

For general $k$, one may extend these results to construct a 
family of nilpotent supersymmetry operators with adjustable parameters
$x_1,\ldots,x_{k-1}$. This construction goes as follows.
Define operators $d_{[a,b],i}$, $d^\dagger_{[a,b],i}$, 
with $a,b=1,\ldots,k$, $b\leq a$, and with $d_{[a,b],i}$ taking out a 
fermion at a site $i$, 
which is at the $b$-th position of a length-$a$ string of consecutive 
occupied sites. The definition of $d_{[a,b],i}$ includes the appropriate
minus sign for bringing the annihilation operator to site $i$. 
Weighing the contribution of $d_{[a,b],i}$ to the supersymmetry 
generator with $y_{a,b}$, one arrives at 
\be
Q^- = \sum_i \sum_{a,b} y_{a,b} d_{[a,b],i} \ .
\ee
Imposing then that $(Q^-)^2=0$ leads to recursion relations on the
coefficients $y_{a,b}$, which can be solved in a closed form 
that is reminiscent of Newton's binomium. Writing $y_a=y_{a,1}$,
the solution reads
\be
y_{a,b} = \frac{y_a y_{a-1} \ldots y_{a-b+1}}{
           y_{b-1} y_{b-2} \ldots y_1} \ .
\ee 
This leaves ${y_1,y_2,\ldots,y_k}$ as free, adjustable 
parameters in the definition of a nilpotent supersymmetry operator 
on the space ${\cal H}_k$. The resulting Hamiltonian 
depends non-trivially on all $k-1$ ratios $x_i \equiv y_i/y_k$,
$i=1,2,\ldots,k-1$.

\subsection{Witten index and ground-state degeneracy}
\label{sec:WittenIndex}

The algebraic structure 
(\ref{eq:qsquared}), (\ref{eq:hamiltonian}) and (\ref{eq:comH}) 
implies the following simple properties of the spectrum of $H$ \cite{Witten}: 
\begin{itemize}
\item
all energies satisfy $E\geq 0$,
\item
all states with $E>0$ can be organized into doublets
of the supersymmetry algebra,
\item
states are annihilated by both $Q^+$ and $Q^-$ if and
only if they are zero-energy eigenstates of $H$.
\end{itemize}
The Witten index of a 
supersymmetric theory is
\be
W_k = \hbox{Tr}_{{\cal H}_k} (-1)^F \exp(- \beta H) \ .
\ee
Because all states with $E>0$ appear in doublets with the same
energy and fermion number differing by $1$, their contribution
to $W_k$ cancels. Thus $W_k$ counts the number of ground states
of the theory, weighted by $(-1)^F$.
A non-zero Witten index requires the existence of at least $|W_k|$
ground states with $E=0$. Note that $W_k$ is independent of
$\beta$, and in fact $H$ itself, as long as supersymmetry
is preserved. The number of ground states in the theory can change
as the parameters change, but $W_k$ cannot. This means ground
states can only appear or disappear in pairs with opposite values
of $(-1)^F$.

Since $W_k$ is independent of $\beta$, we can evaluate it at 
$\beta\to 0$, so that all states contribute 
with a factor $(-1)^F$. 
We compute this by employing a
recursion matrix to pass from $N$ to $N+1$ sites, while
keeping track of the configuration near one open
end. The characteristic equation for the matrix
pertaining to the space ${\cal H}_k$ is
$$
\lambda^{k+1} -\lambda^{k} + \ldots - (-1)^k = 0 \ ,
$$
which implies that $\lambda^{k+2}=(-1)^k$.  This
means that the Witten index $W_k(N)$ for an $N$-site chain
obeys $W_k(N+k+2)= (-1)^k W_k(N)$.
We obtain (for $N=1,\ldots k+2$)
\begin{itemize}
\item
${\cal H}_k$ on an $N$-site open chain,
\begin{equation}
W_k(N) = 
\begin{cases}
0 &{\rm for}\ 1 \leq N \leq k \\
(-1)^k &{\rm for}\ N= k+1,k+2 
\end{cases}
\end{equation}
\item
for ${\cal H}_k$ on an $N$-site closed chain with periodic boundary
conditions,
\begin{equation}
W_k(N) =
\begin{cases} (-1)^{N+1}  &{\rm for}\ 1 \leq N \leq k+1 
\\
 (-1)^k (k+1)  &{\rm for}\ N= k+2 
\end{cases}
\end{equation}
\end{itemize}

For the case with periodic boundary conditions, one can refine
this result by restricting the index
to sectors with a specific eigenvalue $t$ of the shift 
operator $T$. As $T^N=1$, these eigenvalues are of the 
form $t=\exp({2 \pi{\rm i} l /N} )$, $l=1,2,\ldots,N$.
For the case $N=n(k+2)$, where $|W|=k+1$ we find
\be
W=(-1)^N \quad
{\rm for}\  t= (-1)^{N+1} \exp\left(\frac{2\pi{\rm i}}{ k+2}\, j \right)
\qquad{\rm with}\ j=1,\ldots,k+1
\ee
and $W=0$ otherwise. For all other $N$
\be
W=(-1)^{N+1} \qquad {\rm at}\ t=(-1)^{N+1} \ ,
\ee
and $W=0$ otherwise.

These values of the index are independent of the $x_i$: as noted
above, Witten's index is independent of all supersymmetry-preserving
deformations of the theory.  We will show in subsequent sections that
in two simple cases, $M_1$ and $M_2[x=0]$, the eigenstates of
$H$ can be found using the Bethe ansatz. In these two cases, and
for $M_2[x=\sqrt{2}]$ we also
have non-trivial mappings relating the supersymmetric chain to other
integrable models.  The models $M_k[x_1\dots]$ do not appear to be
integrable in general, however.

\section{$M_1$: a supersymmetric model of hard-core fermions}
\label{sec:M1}

\subsection{Hamiltonian}

The Hamiltonian for the $k=1$ model is found by working out
the general expression eq.~(\ref{eq:hamiltonian}) for the
supersymmetry generators defined in eq.~(\ref{eq:Qpm1}). Denoting
$d_i\equiv {\cal P}_{i-1} c_i {\cal P}_{i+1}$ to be the hard-core
fermion, we have
\begin{equation}
H_1 = \sum_{i=1}^N {\cal P}_{i-1} {\cal P}_{i+1} + 
d^\dagger_{i+1}d_i + d^\dagger_{i}d_{i+1}.  
\end{equation}
Henceforth we take periodic boundary conditions, so the index $i$ is
defined mod $N$. 
The potential counts the number of ``empty pairs'', vacancies two 
sites apart. Adding a fermion more than two sites from any other 
fermions decreases the potential by 2, since there are now two less 
empty pairs. It is convenient to rewrite the Hamiltonian as
\begin{equation}
H_1 = N-2f + \sum_{i=1}^N \left[ d_{i+1}^\dagger d^{ }_i + d_i^\dagger d^{ }_
{i+1}+d_i^\dagger d^{ }_{i} d_{i+2}^\dagger d^{ }_{i+2} \right].
\label{fham}
\end{equation}
In this formulation, the Hamiltonian has a constant term, a chemical
potential of $2$ per fermion, a hopping term, and a repulsive potential
for fermions two sites apart. This looks like a
lattice version of the Thirring model with a hard-core
repulsion, and a next-nearest-neighbor repulsive potential.

We remark that the Hamiltonian (\ref{fham}) has been introduced and
analyzed in the literature; it is a special case of the so-called 
constrained XXZ model discussed in \cite{AB-KO}. These papers do not 
mention the supersymmetry properties which form the focus of our analysis 
here.

\subsection{The correspondence with the XXZ model}
\label{sec:m1xxz}

The Hamiltonian (\ref{fham}) bears a simple
relation with the XXZ Hamiltonian with $\Delta=-1/2$
\begin{equation}
H_{{\rm XXZ}-\frac{1}{2}} = {\scriptstyle \frac{1}{2}}
\sum_{j=1}^L \left[\sigma^{\rm x}_j\sigma^{\rm x}_{j+1} +
\sigma^{\rm y}_j\sigma^{\rm y}_{j+1} -
\Delta \sigma^{\rm z}_j\sigma^{\rm z}_{j+1}\right]
\label{xxzh}
\end{equation}
Here the $\sigma$ are Pauli spin matrices. The transformation can be
effected on a periodic chain as follows. Each edge between empty sites
is replaced by an up-spin, and each occupied site with its adjacent
edges is replaced by a down-spin.
The length $L$ of the XXZ chain is therefore related to the length of
the fermion lattice as $L=N-f$.
To transform the Hamiltonian, observe that 
the hopping term simply allows a fermion to move to a neighboring
site, provided its other neighbor is empty.  The effect is the same as
that of the terms $1/2( \sigma^{\rm x}_j\sigma^{\rm x}_{j+1} +
\sigma^{\rm y}_j\sigma^{\rm y}_{j+1})$ in the XXZ Hamiltonian.
The diagonal term in (\ref{fham}) counts the number of empty
second-neighbor pairs and translates into the operator $\sum_j
(1-\sigma^{\rm z}_j)/2 + (1+\sigma^{\rm z}_j)(1+\sigma^{\rm
z}_{j+1})/4$, which counts the number of down-spins and the number of
nearest neighbor up-spin pairs.  On a periodic chain this expression is
identical to $(3L + \sum_j \sigma^{\rm z}_j\sigma^{\rm z}_{j+1})/4$.
This gives, up to an additive constant, a map between the Hamiltonians, 
$H_1 \leftrightarrow H_{{\rm XXZ}-1/2} +3L/4$. Note, however,
that the constant $3L/4$ in the spin chain would not be constant in the
fermion model, as $L=N-f$.

In spite of this reasoning these Hamiltonians 
do not have entirely the same spectrum. One reason for this is that the
size of the Hilbert space is not the same between the two models, even
in the corresponding sector. The size of the sector $(N,f)$ in the
fermion chain is (for $0<f<N/2$ and $N/f$ not an integer)
$$N\left( {\scriptstyle\begin{array}{c}
N-f-1\\ f-1\end{array}}\right),$$
while the size of the corresponding sector in the XXZ chain is
$$\left(\begin{array}{c}N-f\\f\end{array}\right).$$
The second difference is the fact that if a fermion on 
position $N$ hops to the right to position $1$, the weight of the
resulting state will have an additional minus sign if the total number
of fermions is even.

To relate the spectra of the two models 
we first identify the missing states of the fermion chain.
The rules given above give a one-to-one correspondence 
with states of the XXZ chain, for 
all states in the fermion chain in which the leftmost position is empty.
The action of the Hamiltonian (\ref{fham}) within this subspace is well
represented by the expression (\ref{xxzh}), but of course
the Hamiltonian will also hop particles onto the leftmost site.
In order to adjust the Hamiltonian (\ref{xxzh}) to account for jumps outside
the subspace, we have to restrict it to sectors where the translation
operator $T$ has eigenvalue $t$.  
Within this sector each state can be identified with $t$ ($t^{-1}$)
times its left (right) translated version. 
When in the fermion chain the
leftmost position is occupied by a left hop, we identify this state with
$t$ times its left translate. With this
identification, the hop corresponds precisely with a spin exchange 
between the leftmost and the rightmost position, leaving the rest of the
configuration unaffected.
If the number of fermions is even an additional minus sign should be
taken into account for the cyclic permutation of the fermions in this
move.  The result can be emulated in the in the XXZ Hamiltonian by a
twist in the periodic boundary conditions: $\sigma^+_{L+1} =
(-1)^{f+1} t\sigma^+_1$ and $\sigma^-_{L+1} = (-1)^{f+1}
t^{-1}\sigma^-_1$.  Notice that with this boundary condition the
eigenvalues of $T$ in the spin model are not the usual roots of unity
$t^L=1$, but those of the corresponding fermion chain $t^N=1$.  With
this modification the spectrum of the XXZ Hamiltonian in the
appropriate momentum sector corresponds precisely to that of the
fermionic chain. We will confirm this result below by showing that the
Bethe-ansatz equations of the fermionic chain of $N$ sites are the
same as the XXZ chain of length $L=N-f$ if one includes a twist factor
$t (-1)^{f+1}$ in the latter.

\subsection{The Bethe ansatz}

\subsubsection{The Bethe equations}

The Bethe ansatz computation follows closely the original 
computation of Bethe for the Heisenberg spin-chain.
For a review, see {\it e.g.}\ \cite{Baxter,Lowenstein}.
An eigenstate with $f$ hard-core fermions is of the form
\begin{equation}
\phi^{(f)} = \sum_{\{i_j\}} \varphi(i_1,i_2,\dots i_f)
d^\dagger_{i_1}d^\dagger_{i_2}\dots d^\dagger_{i_f}|0\rangle
\label{phif}
\end{equation}
where we order $1\le i_1<i_2-1 <i_3-2 \dots$. The state
$|0\rangle$ is the vacuum state with no fermions, not the ground state.
First we construct eigenstates of the translation operator $T$, which
sends $T d^\dagger_{i_1} d^\dagger_{i_2}\dots = d^\dagger_{i_1+1} d^\dagger_{i_2+1}\dots$.
If there is no fermion at site $N$, then this can be done simply by taking
$$\varphi(i_1,i_2,\dots i_f) \sim
\mu_1^{i_1}\mu_2^{i_2}\dots \mu_f^{i_f}.$$
but this prescription does not work if $i_f=N$.
The way to fix this is to sum over permutations
of the $\mu_i$. Denoting $P$ to be the permutation $(P1,P2,\dots,Pf)$
of $(1,2,\dots f)$, we define
\begin{equation}\varphi(i_1,i_2,\dots i_f) = \sum_P
A_P\,  \mu_{P1}^{i_1}\mu_{P2}^{i_2}\dots \mu_{Pf}^{i_f}.
\label{phidef}
\end{equation}
The periodic boundary conditions give a constraint on
the amplitudes $A_P$. 
For $\phi^{(f)}$ to be an eigenstate of $T$, 
the amplitudes must be cyclically
related as
\begin{equation}
 A_{Pf,P1,P2,\dots P(f-1)} = \mu_{Pf}^N 
 (-1)^{f-1} A_{P1,P2,\dots Pf}.
\label{Acyclic}
\end{equation}
The eigenvalue $t$ of $T$ is then given by
\begin{equation}
t = \prod_{i=1}^f (\mu_i)^{-1}
\label{ev}
\end{equation}
For periodic boundary conditions, untwisted
in the fermionic basis, we set $t^N=1$. 
Defining the bare momentum $p_i$ as $\mu_i\equiv \exp({\rm i} p_i)$,
the relation (\ref{Acyclic}) corresponds to
quantizing the momentum of a particle in a box.

Equation (\ref{phidef}) is Bethe's ansatz for the eigenstates.  
There is no {\it a priori} 
guarantee that it will work, but in this case it does.
The $\mu_i$ and the amplitudes $A_P$ 
are found by demanding that $\phi^{(f)}$ be an
eigenstate.
Let us first study the case with two fermions. We have
$$\phi^{(2)} = \sum_{i_1<i_2-1} 
\varphi(i_1,i_2) d^\dagger_{i_1} d^\dagger_{i_2}
$$
where the sum on $i_1$ and $i_2$ is restricted to
be $1\le i_{1}<i_{2}-1\le N-1$. The Bethe ansatz is that
\begin{eqnarray*}
\varphi(i_1,i_2)&=&\sum_P A_P\  \mu_{P1}^{i_1}\mu_{P2}^{i_2}\\
&=&A_{12} \mu_{1}^{i_1}\mu_{2}^{i_2} 
+A_{21} \mu_{2}^{i_1}\mu_{1}^{i_2}
\end{eqnarray*}
Requiring $\phi^{(2)}$ be an eigenstate of $T$ means that
$$A_{12}=-A_{21} \mu_1^N,$$
with $t=(\mu_1\mu_2)^{-1}$.
Operating with the Hamiltonian (\ref{fham})
on $\phi^{(f)}$, one finds
\begin{eqnarray}
\nonumber
H\phi^{(f)} &=& (N-2f)\phi^{(f)}\  +\  
\sum_{i=1}^N \varphi(i,i+2)\, d^\dagger_{i} d^\dagger_{i+2}
\\
&&\qquad +\sum_{i_1<i_2-1} \varphi(i_1,i_2) \left[
d^\dagger_{i_1+1} d^\dagger_{i_2} 
+ d^\dagger_{i_1-1} d^\dagger_{i_2} + 
d^\dagger_{i_1} d^\dagger_{i_2+1} + 
d^\dagger_{i_1} d^\dagger_{i_2-1}\right] \ .
\label{Hphi}
\end{eqnarray}
Shifting the latter sums by
$i_1\to i_1-1,\ i_1\to i_1+1, 
i_2\to i_2-1$ and $i_2\to i_2+1$, respectively,  yields
$$H\phi^{(f)}=E\phi^{(f)} + 
\sum_{i=1}^N X_i d^\dagger_{i} d^\dagger_{i+2}$$
where
$$E=N-2f + \mu_1 + \mu_2 + (\mu_1)^{-1} + (\mu_2)^{-1}.$$
The other terms must vanish for $\phi^{(f)}$ to be an eigenstate of
$H$. They involve fermions two sites apart, and are
$$X_i = \varphi(i,i+2) -\varphi(i,i+1) -\varphi(i+1,i+2).$$
The first contribution to $X_i$ comes from the potential energy.
The other two are a little bit more subtle.  
Because of the hard-core repulsion, there are no terms in
(\ref{Hphi}) of the form $\varphi(i,i+1)d^\dagger_{i}d^\dagger_{i+1}$. 
After the shifts in $i_1$ and $i_2$, such terms would result in necessary
contributions to $E\phi^{(f)}$. However, since they are lacking,
we need to add them into $E\phi^{(f)}$ and therefore 
subtract them off from the $X_i$.

The terms $X_i$ must all vanish if $\phi^{(f)}$ is to be an eigenstate:
for all $i$
$$X_i=\sum_P A_P\, \mu_{P1}^{i}\mu_{P2}^{i}
\left(\mu_{P2}^2 - \mu_{P2}^{}-\mu_{P1}^{}\mu_{P2}^2\right)=0$$
Each of these vanishes if
$$\frac{A_{21}}{A_{12}}=-\frac{\mu_2}{\mu_1}
\frac{\mu_{2}\mu_1 +1 - \mu_{2}} {\mu_{2}\mu_1 +1 - \mu_{1}}.$$
Combining this with the earlier condition yields
$$ \frac{A_{12}}{A_{21}}=-\mu_1^N = -\frac{\mu_1(\mu_1\mu_2 +1 -\mu_1)}
{\mu_2(\mu_1\mu_2 +1 -\mu_2)}.$$
Using the fact that $t^{-1} = \mu_1\mu_2$ gives the Bethe equation
$$
\mu_k^{N-2} t^{-1} = \frac{\mu_k\mu_j +1 -\mu_k}
{\mu_k\mu_j +1 -\mu_k},$$
which holds for $(k,j)=(1,2)$ and $(2,1)$. One solves these two
equations for $\mu_1$ and $\mu_2$ subject to the constraint 
$t^{-1} = \mu_1\mu_2$. Then the corresponding eigenstate is found
(up to an overall constant) by substituting the values of $\mu_i$
into the equation for $A_{21}/A_{12}$.
In general, one 
must also restrict $\mu_1\ne \mu_2$, or else $\phi^{(2)}$ vanishes. There
is one exception to this: if $\mu_1=\mu_2=\exp(\pm {\rm i}\pi/3)$,
then the $X_i$ vanish without imposing a condition on the $A_P$.
These sort of states are discussed in detail in \cite{BaxBarry},
where it is explained how the Bethe ansatz gives a complete set of
states.
For example, one can easily check that $\mu_1=\mu_2=\exp(\pm {\rm i}\pi/3)$ yields
eigenstates of energy $E=N-2$ when $N$ is an odd multiple of $3$.

This computation can be generalized to all $f$.
The constraint that $\phi^{(f)}$ be an eigenstate is basically
the same as the two-fermion case: one requires that
\begin{equation}
\varphi(i_1,i_2, \dots i,i+2, \dots, i_f) -
\varphi(i_1,i_2, \dots, i,i+1, \dots, i_f) 
-\varphi(i_1,i_2,\dots,i+1,i+2,\dots, i_f)=0
\label{phieigen}
\end{equation}
for any choice of $i_1,i_2,\dots i_f$ and $i$. 
The trick to make this vanish is to consider the permutation
$P'$, which differs from $P$ only in that $Pk$ and $P(k+1)$ are reversed,
i.e.\ $P'=P1P2\dots P(k-1) P(k+1) Pk P(k+2)\dots Pf$.
Then (\ref{phieigen}) can be written as
\begin{eqnarray*}
0&=\sum_{P}^{} \mu_{P1}^{i_1}\mu_{P2}^{i_2}\dots \mu_{Pk}^{i}
\mu_{P(k+1)}^{i}\dots\mu_{Pf}^{i_f}
\Big[A_P\left(\mu_{P(k+1)}^2 - \mu_{P(k+1)}^{} -\mu_{Pk}^{}
\mu_{P(k+1)}^2\right)&
\\
&\qquad\qquad
+A_{P'}\left(\mu_{Pk}^2 - \mu_{Pk}^{} -\mu_{P(k+1)}^{}\mu_{Pk}^2\right)\Big]&
\end{eqnarray*}
This vanishes if for all $P$ and $k$
\begin{equation}
\frac{A_{P'}}{A_P} = g(\mu_{P(k+1)},\mu_{Pk})
\label{AA}
\end{equation}
with 
$$g(a,b)\equiv - \frac{a(ab+1-a)}{b(ab+1-b)}.$$
One can think of $g$ as the bare S-matrix describing
the phase shift when two fermions are interchanged.
The fact that $g(a,a)=-1$  means
that the wavefunction vanishes if two of the $\mu_j$ are identical
(except for $\mu_j=\exp(\pm {\rm i}\pi/3)$, as discussed above).

To find the $\mu_i$, we impose the boundary condition (\ref{Acyclic}).
Note that
\begin{eqnarray*}
\frac{A_{Pf,P1,P2,\dots P(f-1)}}
{A_{P1,P2,\dots Pf}} &=&
\frac{A_{P1,Pf,P2,\dots P(f-1)}}
{A_{P1,P2\dots Pf}} g(\mu_{Pf},\mu_{P1})\\
&=& \prod_{j=1}^{f} g(\mu_{Pf},\mu_{Pj})
\end{eqnarray*}
The condition (\ref{Acyclic}) must hold for all $Pf$, and hence all $j$.
Putting this all together yields the Bethe equations for the $\mu_j$:
$$\mu_j ^N = (-1)^{f} \prod_{k=1}^f g(\mu_j,\mu_k).$$
This is thus a coupled set of polynomial equations for the $\mu_i$.
Using the explicit form of $g$, and the expression (\ref{ev}) for
the translation eigenvalue $t$, the Bethe equations simplify to
\begin{equation}
\mu_j^{N-f} t^{-1} = \prod_{k=1}^f \frac{\mu_j\mu_k + 1 -\mu_j}
{\mu_j\mu_k +1 -\mu_k}.
\label{bethe}
\end{equation}
These are a set of $f$ coupled polynomial equations for the
$\mu_j$, $j=1\dots f$. Solving these for a set of $\mu_j$, one then finds the
corresponding
eigenstate (up to an overall normalization) by using (\ref{AA}).
This eigenstate has energy
\begin{equation}
E= N-2f + \sum_{j=1}^f \left[\mu_j + (\mu_j)^{-1}\right]
\label{energy}
\end{equation}

The correspondence to the XXZ model at $\Delta =- 1/2$ discussed above
is readily apparent in the Bethe equations (\ref{bethe}).
These are precisely the Bethe equations one obtains for this XXZ
model with $L=N-f$ sites, and with a twist resulting in the factor of $t^{-1}$
in (\ref{bethe}). However, the eigenstates in
the two cases are slightly different: the factor $\mu_{P(k+1)}/\mu_{Pk}$
in (\ref{AA}) does not appear in the XXZ case.

In the preceding Bethe ansatz computation, we have not used the
supersymmetry at all.  In fact, one can easily repeat the computation
for a model with an arbitrary coefficient of the four-fermion term in
(\ref{fham}).  Such a model is integrable (it is related to the XXZ
chain at arbitrary $\Delta$), but it is no longer supersymmetric. It
would be interesting to understand if there is a quantum-group
symmetry generalizing the supersymmetry to arbitrary four-fermion
coupling. In particular, it is plausible that the loop group symmetry
of \cite{barry} specialized to $\Delta=-1/2$ is related to the supersymmetry
described here.

\subsubsection{The monomer Bethe ansatz}

The Bethe ansatz has been argued to be complete \cite{BaxBarry}.
However, when $f>N/3$, there seem to be too many
solutions to the Bethe equations. For example, consider the state with
the maximum number of fermions, which has $f=N/2$ when $N$ is even. In
this state, the hard-core fermions (or equivalently, the hard-core
dimers) are closely packed.  There are just two eigenstates for this
value of $f$, but the Bethe
equations are suggestive of many more solutions. Since however, all
solutions to the Bethe equations in which the $\mu$ are either distinct
or equal to  $\exp(\pm {\rm i} \pi/3)$ give an eigenvector, the number of
solutions can not exceed the number of eigenvectors unless dependent
vectors are constructed.
The analogous situation in the XXZ model is easy to
understand: the two situations are related by an overall spin flip,
and the corresponding solutions to the Bethe equations can be related.
Because of the hard-core constraint in our problem, the particle/hole
transformation is trickier;  it can be found using the
mapping to the XXZ model described above, but will relate systems with
different numbers of sites. 
To understand the states with $f>N/3$, it is thus
useful to build up a Bethe ansatz solution
starting from the close-packed state. It will turn out
as well that the resulting Bethe ansatz equations can also be defined
even in the case that the states have fractional charge.

For $N$ even, there are two
close-packed states. In terms of fermions they can be written
as
$$|C_1\rangle = d_1^\dagger d_3^\dagger\dots d_{N-1}^\dagger|0\rangle
\qquad\qquad |C_2\rangle = d_2^\dagger d_4^\dagger\dots d_{N}^\dagger|0\rangle
$$
States with smaller $f$ are described
by having domain walls between the two reference states.
A state with $f=N/2 -1$ has two domain walls.
It is convenient to pictorially describe each hard-core 
fermion as a dimer centered on site $i$. The domain walls then correspond
to monomers: sites on the dual lattice unoccupied by a dimer.
The two reference states and an example of a state with
two monomers are then 
\newcommand{\dimermono}{\rule[.035in]{.27in}{.015in}
\hspace{.225in}\circ\hspace{.20in}}
\newcommand{\dimer}{\rule[.035in]{.27in}{.015in}\hspace{.28in}}
\
\begin{eqnarray*}
|C_1\rangle &=&\dimer  \dimer  \dimer  \dimer 
\dimer  \dimer \dimer \dots \\
|C_2\rangle &=&\hspace{.28in} \dimer  \dimer  \dimer  
\dimer  \dimer  \dimer\quad \dots\\
m_5^\dagger m_{10}^\dagger|C\rangle &=&
\dimer  \dimermono \dimer  \dimermono
\dimer  \dimer \dots
\end{eqnarray*}
We have denoted by $m_i$ the operator which creates a 
monomer at site $i$ on the dual lattice. 
These of course can be written
in terms of the original dimer creation operators: for example,
$m_5^\dagger m_{10}^\dagger|C\rangle = d_1^\dagger
d_3^\dagger d_6^\dagger d_8^\dagger  d_{11}^\dagger 
 d_{13}^\dagger\dots |0\rangle$.

All of the states in the Hilbert space can be described in this monomer
basis. For $N$ even, there must be an even number of monomers $M$ to
ensure the periodic boundary conditions.  The monomer construction
works for odd $N$ as well: such states correspond to odd $M$. 
For both even and odd $N$, the fermion number $f$ is 
$f=(N-M)/2$.  Monomers can be placed next to each
other; two adjacent monomers simply correspond to removing a dimer.
The only restriction on monomer placement is that sequential monomers
must be an odd number of sites apart.  We can thus write any state as
$$\phi^{(f)} = \varphi_M(i_1,i_2,\dots,i_M) 
m^\dagger_{i_1}m^\dagger_{i_2}\dots m^\dagger_{i_M}|C\rangle.$$
with $1\le i_1<i_2<\dots <i_M \le N$, and
$i_{j+1}-i_j$ must be an odd number.
There is no necessity for the subscript of $|C\rangle$
when there are monomers present: the monomer creation operators
are sufficient to specify the state uniquely.

The Hamiltonian in the monomer basis is
$$H_M = \frac{N}{2} -\frac{M}{2} + 
\sum_{i=1}^N  \left[\left( m^\dagger_{i+2} m_{i}^{ } 
+ m^\dagger_{i} m_{i+2}^{ }\right)
\left(1- m^\dagger_{i+1} m_{i+1}^{ }\right) + m^\dagger_{i+1} m_{i+1}^{ }
m^\dagger_{i} m_{i}^{ }\right]$$
A monomer can hop two sites as long as it does not hop
over another monomer; this preserves the restriction that $i_j-i_{j+1}$
be odd.
As for dimers, the potential energy goes up when monomers are adjacent.

The one subtlety in working in the monomer basis is the implementation
of the translation operator. The translation operator takes
$T|C_1\rangle = |C_2\rangle$, but because 
the dimer creation operators anticommute, 
$T|C_2\rangle = -(-1)^{f}|C_1\rangle$. In general,
one picks up this minus sign if $f$ is even and if
the first monomer is on an even site.
It is thus convenient to definite some of the
monomer states with an extra factor of $-{\rm i}$, namely 
$m^\dagger_{i_1}m^\dagger_{i_2}\dots|C\rangle\equiv
-{\rm i}\, d_2^\dagger d_4^\dagger\dots|0\rangle$
when both $f$ and $i_1$ are even. With this definition, one
has for states with no monomer on the last site $N$,
\begin{eqnarray*} 
T(m^\dagger_{i_1}m^\dagger_{i_2}\dots|C\rangle)= \lambda
m^\dagger_{i_1+1}m^\dagger_{i_2+1}\dots|C\rangle
\end{eqnarray*}
where $\lambda = {\rm i}$ if $f$ is even, and $\lambda=1$ if $f$ is odd.
If there is a monomer on the last site $N$,
one has (for any $f$)
$$T(m^\dagger_{i_1}m^\dagger_{i_2}\dots m_N^\dagger |C\rangle)=
m_1^\dagger m^\dagger_{i_1+1}m^\dagger_{i_2+1}\dots m^\dagger_{i_{M-1}+1}
|C\rangle.$$
The latter relation means that the monomer operators commute with
each other.

The Bethe ansatz analysis is very
similar to that in
the original basis. We take
\begin{equation}\varphi_M(i_1,i_2,\dots i_M) = \sum_P
B_P\  \mu_{P1}^{i_1}\mu_{P2}^{i_2}\dots \mu_{Pf}^{i_f}.
\label{phidef2}
\end{equation}
The eigenvalue of the translation operator $T$ is then
$$t=\lambda \prod_{j=1}^M (\mu_j)^{-1}.$$
Imposing the periodic boundary conditions yields
\begin{equation}
\lambda B_{PM,P1,P2,\dots P(f-1)} = \mu_{Pf}^N B_{P1,P2,\dots PM}.
\label{Bcyclic}
\end{equation}
For (\ref{phidef2}) to describe an eigenstate of the Hamiltonian,
the amplitudes $B_P$ must obey
\begin{equation}
\frac{B_{P'}}{B_P}=-\frac{\mu_{Pk}}{\mu_{P(k+1)}}
\frac{\mu_{Pk}^2\mu_{P(k+1)}^2+1 - \mu_{P(k+1)}^2}
{\mu_{Pk}^2\mu_{P(k+1)}^2+1 - \mu_{Pk}^2}
\label{BB}
\end{equation}
where the permutation $P'$ is defined as before.
Combining (\ref{BB}) with
the cyclic relation (\ref{Bcyclic}) yields the Bethe equations. 
In terms of new variables $\nu_j\equiv(\mu_j)^2$,
they resemble those in the original basis; they are
\begin{eqnarray}
\nu_j^{N-f} t &=& (-1)^{M+f}\prod_{k=1}^M \frac{\nu_j\nu_k + 1 -\nu_j}
{\nu_j\nu_k +1 -\nu_k}\\
\prod_{j=1}^M \nu_j&=& (-1)^{f+1} t^{-2}
\label{betheM}
\end{eqnarray}
where we used the fact that $f=(N-M)/2$.
The energy of this eigenstate is
\begin{equation}
E= f + \sum_{j=1}^M \left[\nu_j + (\nu_j)^{-1}\right]
\label{energyM}
\end{equation}
The similarity of the Bethe equations in the two bases
is of course a consequence of the mapping to the XXZ
model; in the XXZ model there is an obvious symmetry between up and
down spins. 

\subsubsection{Supersymmetry and the ground state}
\label{sec:susyGS}

In this section we describe several interesting consequences
of the supersymmetry for the Bethe ansatz.

The simplest one is the pairing of
eigenstates of $H$.
Acting with $Q^+$ on the state $\phi^{(f)}$ defined in (\ref{phif}),
we have
$$Q^+\phi^{(f)} = \sum_{\{i_j\}} \varphi(i_1,i_2,\dots i_f)
\left(
\sum_{i=1}^f d^\dagger_i d^\dagger_{i_1}d^\dagger_{i_2}\dots d^\dagger_{i_f} 
|0\rangle\right).$$
This is a state with $f+1$ fermions, and so can also be written
in the form (\ref{phif}) where $\varphi$ takes a special form
$\varphi^+$, namely
$$\varphi^+ = \sum_{i=1}^{i_1-2}\varphi(i,i_1,\dots i_f) - 
\sum_{i=i_1+2}^{i_2-2}\varphi(i_1,i,i_2\dots i_f) + \dots + (-1)^{f+1}
\sum_{i=i_f+2}^N\varphi(i_1,i_2,\dots,i_f,i).$$
The minus signs are a result of the fermions anticommuting.
If the state $\phi^{(f)}$ is of
Bethe ansatz form (\ref{phidef}), the 
state $Q^+\phi^{(f)}$ is also of Bethe ansatz form, with
$$\varphi^+ = \varphi(1,\mu_1,\mu_2,\dots\mu_f).$$ The relative minus
signs work out correctly from (\ref{AA}) because $g(1,\mu_j)=-1$. It
is easy to check that if a state $\phi^{(f)}$ obeys the Bethe
equations (\ref{bethe}) with a set $\mu_1,\mu_2,\dots,\mu_f$, then the
set $(1,\mu_1,\mu_2,\dots,\mu_f)$ also obeys the Bethe
equations. Moreover, the two states have the same energy.

We have thus shown that acting with the supersymmetry generator $Q^+$
corresponds to inserting a particle with zero bare momentum, where the
bare momentum $p_j$ is defined via $\mu_j\equiv \e^{{\rm i} p_j}$.  Acting
with $Q^+$ twice annihilates the state, because states with $\mu_j=
\mu_k=1$ for $j\ne k$ have vanishing wavefunction. 
Thus $(Q^+)^2=0$ as required.  To complete
this identification of doublets within the Bethe ansatz one can verify
explicitly that any state obeying the Bethe equations satisfies
$Q^{-}\phi^{(f)} =0$ unless one of the $\mu_j=1$. An $E=0$ ground
state $\phi_0$ therefore has all $\mu_j\ne 1$; it also has the special
property that $Q^+\phi_0=0$ even though none of the $\mu_j$ are $1$.

An even deeper consequence of the supersymmetry results from the fact
that the ground state energy is zero.  We will derive a single
polynomial whose roots give the roots of the Bethe equations $\mu_j$
(or $\nu_j$) describing the ground state.  Thus for the ground state
we reduce a set of $f$ coupled polynomial equations of order $f$ to a
single one of order $f$.  Results very similar to those in this
section have been derived for the XXZ chain at $\Delta=-1/2$ for an
odd number of sites, or for special twisted boundary conditions
\cite{delta-half}. The computations of \cite{delta-half} were possible
because the transfer-matrix eigenvalues of the corresponding
six-vertex model were known explicitly, through some clever arguments
\cite{Baxter}. Our result thus explains why these eigenvalues could be
found explicitly: in our formulation, the model is supersymmetric, and
so the ground state energy vanishes.

To do this computation, we need to make some definitions.
We define $w_j$, $j=1\dots f$, via
$$\mu_j = \frac{w_j-q}{qw_j-1}$$
where $q\equiv\exp(-{\rm i}\pi/3)$. 
In terms of $w$, the energy of an eigenstate is
\begin{equation}
E= N-\sum_{j=1}^f\frac{(w_j+1)^2}{(w_j-q)(w_j-q^{-1})}.
\label{Ew}
\end{equation}
Baxter's ${\cal Q}$-function \cite{Baxter} is
a polynomial with zeroes at $w=w_k$, namely
$${\cal Q}(w)=\prod_{i=1}^f (w-w_k).$$
The Bethe equations can then be written in terms of the $w_j$ as
$$\left(\frac{w_j-q}{qw_j-1}\right)^{N-f} t^{-1}=
q^{-f}\frac{{\cal Q}(q^{-2}w_j)}{{\cal Q}(q^{2}w_j)}
$$
Baxter defined this function for integrable lattice models because
the transfer-matrix eigenvalues can be easily expressed in
terms of it. We define an analogous function ${\cal T}(w)$
by the equation 
\begin{equation}
{\cal T}(w) {\cal Q}(w) (1+w)^{N-f} q^{-N} t
=   {\cal Q}(q^{-2} w) (q^{-2}w +1)^{N-f}
-  t^{-1} q^{-N}{\cal Q}(q^2 w) (q^2w+1)^{N-f}
\label{TQ}
\end{equation}
Note that ${\cal T}(w_k)$ is finite
even though ${\cal Q}(w_k)=0$, 
because the Bethe equations require the right-hand side 
of this equation to vanish when $w=w_k$ as well. To simplify this relation,
we define ${\cal R}(w)={\cal Q}(w)(1+w)^{N-f}$, so that 
\begin{equation}
{\cal T}(w) {\cal R}(w)q^{-N} t
=   {\cal R}(q^{-2} w)-  t^{-1} q^N {\cal R}(q^2 w) .
\label{TR}
\end{equation}
The eigenvalue $t$ of the translation operator can be expressed as
\begin{equation}
t=\prod_{i=1}^f (\mu_j)^{-1}= \frac{{\cal Q}(q)}{{\cal Q}(q^{-1})}
q^{-f} = \frac{{\cal R}(q)}{{\cal R}(q^{-1})}
q^{-N}
\label{tR}
\end{equation}

We have defined ${\cal T}(w)$ so that we can obtain the energy eigenvalues
(\ref{energy}) by expanding ${\cal T}(w)$ around $w=q$. Because
${\cal R}(q^3)={\cal R}(-1)=0$, the relations (\ref{TR}) and (\ref{tR})
give
$${\cal T}(q) = t^{-1}q^{N}\frac{{\cal R}(q^{-1})}
{{\cal R}(q)} = 1$$
Expanding ${\cal T}(q)$ and ${\cal R}(q)$ in Taylor series,
and substituting into (\ref{TQ}), relates their derivatives as
\begin{equation}
\frac{{\cal T}'(q)}{{\cal T}(q)}=
q^{-2}\frac{{\cal R}'(q^{-1})}{{\cal R}(q^{-1})}
-\frac{{\cal R}'(q)}{{\cal R}(q)}
\label{Tprime}
\end{equation}
Using the definition of ${\cal R}(w)$, we find that
$$\frac{{\cal R}'(w)}{{\cal R}(w)}= \frac{N-f}{1+w} +
\sum_{j=1}^f \frac{1}{w-w_j}.$$
Substituting this into (\ref{Tprime}), and comparing to the
energy (\ref{Ew}), we find that
\begin{equation}
\frac{{\cal T}'(q)}{{\cal T}(q)} = \frac{1}{q^2-1} E
\label{TE}
\end{equation}

In light of the correspondence of our model with the XXZ model
discussed above, it is natural to expect that ${\cal T}(w)$ for
arbitrary $w$ will provide the eigenvalues for some lattice model with
$w$ as an anisotropy parameter. The eigenvectors of the transfer
matrix of the lattice model will be identical to the eigenvectors of
our quantum Hamiltonian.  These eigenvectors will be independent of
$w$, so the lattice model will have commuting transfer matrices and
hence be integrable as well.  The eigenvalues, which do depend on $w$,
are determined by the same Bethe equations as our fermion chain, via
the ${\cal T}$-${\cal Q}$ relation (\ref{TQ}).  One can also work out a dual ${\cal
T}_M$, using the monomer Bethe ansatz.  We will not, however, pursue
this connection with the lattice model any further.

The fact that the energy is simply related to ${\cal T}(q)$ via
(\ref{TE}) allows us to explicitly
find ${\cal Q}(w)$ for the zero-energy ground states.
Because ${\cal T}(q)=1$ and
${\cal T}'(q)\propto E$, this means that we should search for
a ${\cal Q}(w)$ satisfying (\ref{TQ}) with
${\cal T}(w)=1$ for all $w$. Such a solution (denoted ${\cal Q}_0(w)$)
is the ${\cal Q}$-function for a ground state with $E=0$. The roots of the polynomial
${\cal Q}_0(w)$ are the $w_j$ for this ground state.
Denoting the corresponding ${\cal R}(w)$ by ${\cal R}_0(w)$, we require
therefore
\begin{equation}
{\cal R}_0(q^{-2}w) = t q^{-N}{\cal R}_0(w) + 
t^{-1} q^N {\cal R}_0(q^2 w).
\label{RRR}
\end{equation}
If there exists a solution to this equation with
non-vanishing wavefunction, then we have
an $E=0$ state. We will find such solutions explicitly.

The crucial fact to exploit is that
by definition ${\cal Q}_0(w)$ and ${\cal R}_0(w)$ are polynomials of order
$f$ and $N$ respectively. 
We define the coefficients $r_k$ so that
${\cal R}_0(w)=R_f\sum_{k=0}^N (-w)^k r_k$,
where $R_f$ is an unimportant overall coefficient.
Using this expansion
in (\ref{RRR}) gives
$$\left(q^{-2k} -tq^{-N}- t^{-1}q^{N+2k}\right)r_{k}=0$$
for all $k$. Thus either the expression in parentheses vanishes,
or else $r_k$ does.
This requires that every third term in ${\cal R}_0$ must vanish.
A non-vanishing $r_0$ means that
$1=tq^{-N}+t^{-1}q^N$, so for $N=3p$ with $p$ an integer,
${\cal R}_0\ne 0$ only if $t=(-1)^N q^{\pm 1}$, while 
when $N=3p\pm 1$, one must have $t=(-1)^N$.

Two facts about ${\cal R}_0(w)$ give us enough information to
find it exactly, following the method of \cite{Stroganov}. 
The first is that every third term vanishes.
The second is that ${\cal Q}_0(w)={\cal R}_0(w)/(1+w)^{N-f}$ is a
polynomial of order $f$, so that
$${\cal R}_0(-1)={\cal R}'_0(-1)={\cal R}^{\prime\prime}_0(-1)=
\dots {\cal R}^{(N-f-1)}_0(-1)=0
$$
If $f$ is too small, then there are too many equations to solve,
and no non-trivial solution exists. If $f$ is too large,
one can find solutions, but these must have vanishing
wavefunction (this presumably can be shown by utilizing
the monomer Bethe ansatz).
One finds a non-trivial ${\cal R}_0(w)$ only when $f=f_0\equiv
\hbox{int}((N+1)/3)$.
Thus the ground state has $f=f_0$,
as had already been shown using other techniques in \cite{FSd}.
With a little more work, we find ${\cal R}_0$ explicitly. The non-vanishing
$r_k$ are
\begin{eqnarray*}
r_{N-3k} = (-1)^f r_{3k} &=& B^{-1/3,1/3}_{0,1/3}(k)
\qquad \qquad\ N=3f_0+1\\
r_{N-3k} = (-1)^f r_{3k} &=& B^{1/3,5/3}_{0,2/3}(k)
\qquad\qquad\quad  N=3f_0-1
\end{eqnarray*}
and
$$
r_{3k}=
B^{1/3,2/3}_{1/3,0}(k),
\qquad\qquad
r_{3k+1}=B^{1/3,2/3}_{0,-1/3}(k) \qquad\qquad N=3f_0
$$
where $B^{a,b}_{c,d}(k)$ is the product of binomials
$$
B^{a,b}_{c,d}(k)= 
\begin{pmatrix} f_0-a\\k-c \end{pmatrix}
\begin{pmatrix} f_0-b\\k-d \end{pmatrix}.
$$
There are two ground states for $N=3f_0$;
the other ${\cal R}_0$ is given by exchanging $r_{3k}$ with $r_{N-3k}$.
Dividing by $(1+w)^{N-f}$ gives ${\cal Q}_0$.
There are many
interesting mathematical issues associated with these polynomials. 
In particular, it is from these polynomials that some few of
the very many observed properties of the ground state of the XXZ chain
have now been proven\cite{Gier}.

\subsection{Excitations}

In this subsection, we will show that
the low-energy ``holon'' excitations of this model have fractional charge 
$\pm 1/2$.  
The fractional charge of $1/2$
is quite natural in the monomer basis, because two monomers
have a charge of only one. 
Another hint of charge-$1/2$ particles is the fact 
that there are two ground states when the number of sites is
a multiple of three. We will use the Bethe ansatz to derive
their existence.

The presence of fractional charge was conjectured in \cite{FSd} by
using a heuristic N\'eel-like picture of the ground state. It was
suggested there that excitations with charge $1/3$ appear, but there are
several reasons why these states do not appear in the
spectrum in the continuum limit. The reason for the charge $1/3$ was
that there seem to be three N\'eel-like ground states, differing by
translation by one lattice site. However, the exact index results of
\cite{FSd} and the results for ${\cal Q}_0$ above indicate that there
are either one or two ground states, depending on the number of sites. 
Moreover, in the conformal field theory
describing the continuum limit, the charge-$1/3$ states are created by
a chiral operator called the spinon. 
Although one can build up the Hilbert space of the
conformal field theory by acting with the spinons, 
neither single-spinon nor two-spinon states appear in the spectrum,
as defined by the toroidal partition function.

To find the charge $\pm 1/2$ states,
we need to study the solutions of the Bethe equations
(\ref{bethe}) in the $N\to\infty$ limit. To make contact with
past studies of the XXZ model at $\Delta=-1/2$, it is convenient
to define $u_j=\ln(w_j)/2$ so that 
$$\mu_j\equiv \frac{\sinh(u_j+{\rm i}\pi/6)} {\sinh(u_j-{\rm i}\pi/6)},$$
and the Bethe equations are
\begin{equation}
\left(\frac{\sinh(u_j+{\rm i}\pi/6)} {\sinh(u_j-{\rm i}\pi/6)}\right)^{N-f}
t^{-1}= (-1)^f \prod_{k=1}^f \frac{\sinh(u_j-u_k+{\rm i}\pi/3)} 
{\sinh(u_j-u_k-{\rm i}\pi/3)}.
\label{bethemod}
\end{equation}
In the $N\to\infty$ limit, there are solutions of (\ref{bethemod})
with $u_j$ real, and with $u_j+{\rm i}\pi/2$ real, so that both sides of the
equation have modulus $1$. There are also ``string'' solutions, where
there are two or more $u_j$ with the same real part, and having
imaginary parts differing by $ {\rm i}\pi/3$.  The classic result of
\cite{Taka} is that in the $N\to\infty$ limit, the only string
solution which needs to be included is the $2+$ string, which consists
of pairs $u_j^{(2)}-{\rm i}\pi/6, u_j^{(2)}+{\rm i}\pi/6$, 
with $u_j^{(2)}$ real. Other string
solutions occur at most in finite numbers as $N\to\infty$, and so can
be ignored.

Another long-known result is that the ground state of this model
contains only solutions of (\ref{bethemod}) with real $u_j$ \cite{Baxter}
(to avoid complications involving $1/N$ corrections, we assume
in this subsection that $N$ is a multiple of $3$).
In terms of the bare momenta $p_j$ defined by $\mu_j=\e^{i p_j}$,
real $u_j$ corresponds to $\pi \ge |p_j|> \pi/3$, so one can
think of $p_F = \pm \pi/3$ as the Fermi levels.
One can check using the following analysis that including other
solutions of the Bethe equations raises the energy.
(One can also check that all the
roots of the polynomial ${\cal Q}_0$ correspond to real $u_j$.)

Since there are $f_0 =\hbox{int }[(N+1)/3]$ particles in the ground state,
the number of $\mu_j$ in the ground state also diverges as $N\to\infty$.
We can therefore define a density
$\rho_0(u)$, so that in this limit, the number of solutions of
(\ref{bethemod}) in the ground state with $u_j$ real and in between
$u$ and $u+du$ is $\rho_0(u)du$.
To derive $\rho_0$, we follow the
standard procedure (see {\it e.g.}\ \cite{Lowenstein}). We take the derivative
of the log of (\ref{bethemod}), yielding
$$\rho_0(u) = (N-f_0)\alpha_{1/6}(u) - \int_{-\infty}^\infty
\d u' \rho_0(u')\alpha_{1/3}(u-u'),$$
where
$$\alpha_{a}(u) = -\frac{1}{2\pi{\rm i}}\frac{\partial}{\partial u}
\ln \frac{\sinh(u+{\rm i} a)}{\sinh(u-{\rm i} a)} = \frac{1}{\pi}
\frac{\sin(2a)}{\cosh(2u)-\cos(2a)}$$
This can be solved by Fourier transformation, using the fact that
for $0< a \le 1/2$
$$\widetilde{\alpha}_a(k)=\frac{\sinh[(1-2a)\pi k/2 ]}{\sinh[\pi k/2]}.$$
This yields
$$\widetilde{\rho}_0(k)= \frac{N-f_0}{2\cosh[\pi k/6 ]}, \qquad\qquad
{\rho}_0(u)= \frac{3(N-f_0)}{2\pi\cosh(3u)}.$$
This result can be checked in several ways. First, note that
the fermion number $f_0$ in the ground state is
$$f_0 = \int_{-\infty}^\infty \d u\, \rho_0(u) = \widetilde{\rho}_0(0) =
\frac{N-f_0}{2}.$$
Thus $f_0=N/3$ as shown previously. Second, one can plug in
this expression for $\rho_0(u)$ into (\ref{energy}), and verify
that indeed $E_0=0$.

To understand the excitations over this ground state, we define $P(u)$
to be the density of solutions of (\ref{bethemod}) with $u$ real. For
the ground state with $N$ a multiple of $3$, 
all real solutions are utilized and
$P_0(u)=\rho_0(u)$. For excited states, we allow there to be
holes in this distribution: a hole is a value $u_j$ which solves
(\ref{bethemod}) but which is not one of the $u_k$, $k=1\dots f$. We
denote the values of $u_j$ corresponding to these holes as
$u^{(h)}_a$, $a=1\dots H$.  Excited states can also contain the two
kinds of solutions of (\ref{bethemod}) which have $u$ not real. Those
of the form $u_j=u^{(1)}_b + {\rm i}\pi /2$ with $u^{(1)}_b$ real are
usually called $1-$ strings, and are indexed $b=1\dots {\cal
N}_1$. The $2+$ strings, consisting of pairs $u^{(2)}_c - {\rm i}\pi /6,
u^{(2)}_c + {\rm i}\pi /6$, are indexed $c=1\dots {\cal N}_2$. The Bethe
equation for the Fourier transform $\widetilde{P}(k)$ for the
low-energy excitations is then
\begin{equation}
\widetilde{P}(k) = \widetilde{\rho}_0(k)+\frac{1}{2\cosh[\pi k/6] }
\left(f_0 -f + \sum_{a=1}^H \frac{\widetilde{\alpha}_{1/3}(k)}
{\widetilde{\alpha}_{1/6}(k)} \e^{{\rm i} ku^{(h)}_a}
+\sum_{b=1}^{{\cal N}_1} \e^{{\rm i} ku^{(1)}_b}
-\sum_{c=1}^{{\cal N}_2} \e^{{\rm i} ku^{(2)}_c}\right)
\label{forP}
\end{equation}

We can use (\ref{forP}) to compute the charge of the excitations.
Denoting by ${\cal N}_{\rm real}$ the total number 
of real solutions of the Bethe equations, we have
(\ref{forP}):
$$ {\cal N}_{\rm real}
=\int_{-\infty}^\infty \d u\, P(u) = \widetilde{P}(0) = \frac{3f_0}{2}
-\frac{f}{2}
+ \frac{H}{4} + \frac{{\cal N}_1}{2} - \frac{{\cal N}_2}{2}$$
By definition of $H$, the total number of {\it filled} real solutions is 
${\cal N}_{\rm real} - H$. 
The fermion number $f$ is therefore
$$f={\cal N}_{\rm real} - H + {\cal {\cal N}}_1 + 2{\cal N}_2,$$
the latter factor of $2$ because each $2+$ string corresponds 
to two $\mu_j$
and hence fermion number $2$. 
Thus the fermion number of a low-lying excited state is
$$f=f_0 - \frac{H}{2} + {\cal N}_1 - {\cal N}_2$$ One must choose the
numbers of holes and strings in a given solution so that ${\cal
N}_{\rm real}$ is an integer. This means there must be an even
number of holes, so overall the fermion number is indeed an integer,
as it must be. However, a two-hole state is described by two
variables, $u^{(h)}_1$ and $u^{(h)}_2$. Moreover, one can check that
the energy of this state is the sum of two terms, one depending on
$u^{(h)}_1$ and the other $u^{(h)}_2$.  Thus this state is properly
described as a two-holon state with fermion number $-1$ relative to
the ground state. These holons are identical, so each holon is
interpreted as having fermion number $-1/2$. One can find a charge
$+1/2$ state by acting with the supersymmetry charge $Q^+$; this
yields a state with two holes and one $1-$ string (the latter with
$u^{(1)}= {\rm i}\pi/2$). The state has no net charge over the ground state,
and is interpreted as a holon/antiholon pair.

To find the excitations with positive fractional charge directly, it
is easiest to look at the monomer basis. The analysis is virtually identical
to that above. The number of monomers $M$ is related to fermion number
$f$ by $f=(N-M)/2$.  For these Bethe equations we denote ${\cal
M}_{\rm real}$, ${\cal M}_H$, ${\cal M}_1$, ${\cal M}_2$ and $M_0$ the number of
real solutions, the number of holes, the number of $1-$ strings, the
number of $2+$ strings, and the number of monomers in the ground state. 
We find
$$M= {\cal M}_{\rm real} -{\cal M}_H + {\cal M}_1 + {\cal M}_2$$
and
$$ {\cal M}_{\rm real} = \frac{3M_0}{4} +\frac{M}{4} + 
\frac{{\cal M}_H}{4} + \frac{{\cal M}_1}{2} - \frac{{\cal M}_2}{2}.$$
Putting these two together yields
$$f-f_0 = \frac{M_0-M}{2} = \frac{{\cal M}_H}{2} - {\cal M}_1 + {\cal
M}_2.$$ Thus we find that holes in the sea of monomers have charge
$1/2$.  One slight difference with the preceding analysis is that one
can find solutions of the Bethe equations where ${\cal M}_H=1$, so
that the state seems to have fractional fermion number.  These states
({\it e.g.}\ $N=15$, $M_0=5$, $M=4$, ${\cal M}_{\rm real}=6$) do have ${\cal
M}_{\rm real}$ and $M$ integer. However, these states are not in our
original space of states, because they violate the restriction that
the number of monomers $M$ be odd for odd $N$, and even for even
$N$. However, it is interesting that by extending the space 
of allowed monomer states, one can find sensible Bethe
equations for states with fractional charge.

\subsection{The holons in field theory}
\label{sec:holonsft}

It is not hard to check that the low-energy excitations of the theory
are gapless in the continuum limit. Thus one would expect that there
should be a conformally-invariant field theory describing the model
in this limit. This conformal field theory is
the simplest ${\cal N}$=(2,2) supersymmetric conformal field theory.
The $(2,2)$ means that in this Lorentz-invariant gapless field theory,
the symmetry is enhanced to two left-moving and two right-moving
supersymmetries.
Its relation to this lattice model has been described in detail in
\cite{FSd}. If we assume that the charge $\pm 1/2$ holons form a 
doublet under supersymmetry, this implies that 
the left-moving holon states are $V_{\pm 1/2, \pm 3/4}|0\rangle_{NS}$,
and the right movers are $V_{\pm 1/2, \mp 3/4}|0\rangle_{NS}$. 
The left-moving state has left and right scaling dimensions $(3/8,0)$,
while the right-moving state has $(0,3/8)$. This fits
in nicely if somewhat strangely in the ${\cal N}$=(2,2) language: 
the left-moving holon has left part in the first excited Ramond state,
while the right part is in the Neveu-Schwarz vacuum. Putting
a left and a right holon together gives the dimension $(3/8,3/8)$ state,
which corresponds to a chiral primary field acting on the Ramond vacua.
The holon creation operators acting on
the Ramond ground states (those appearing
in the lattice model with periodic boundary conditions) 
are $V_{\pm 1/2,\pm 1/4}$, and have dimension $5/24$ and Lorentz spin $1/8$.
It would be interesting to use the monomer
Bethe ansatz to compute the finite-size energy of a single holon state 
in order to verify this.

We note also that one can obtain a ${\cal N}$=2 supersymmetric massive
field theory by perturbing the conformal field theory by the only
supersymmetry-preserving relevant operator, $V_{0,2}+V_{0,-2}$, of
dimensions $(2/3,2/3)$.  This model is integrable, and its exact
particle spectrum is known. In the ${\cal N}$=2 language,
this amounts to a doublet of charge $\pm 1/2$ kinks \cite{FI}.
These kinks interpolate between the two vacua of the theory.
It is natural to identify these two fractionally-charged
kinks with the holons at the critical point.

\section{$M_2[x]$: supersymmetric models of single fermions and pairs}

\subsection{The Hamiltonian and the ground states}
\label{sec:HamGS}

Our discussion in section \ref{sec:WittenIndex} 
provides a somewhat implicit definition
of the model $M_2[x]$, with the parameters $y_1$, $y_2$ fixed at
$y_1=x$, $y_2=1$. Its Hamiltonian acts on the space ${\cal H}_2$
of single fermions (s) and nearest neighbor pairs (p) on a 1D lattice,
all spaced by one or more unoccupied sites. The Hamiltonian can be
written out in second quantized form, but we prefer to list the
non-vanishing amplitudes that it entails
\begin{itemize}
\item
single hop, $\ldots 0100 \ldots \leftrightarrow \ldots 0010 \ldots$,
with amplitude $x^2-1$,
\item
pair hop, $\ldots 01100 \ldots \leftrightarrow \ldots 00110 \ldots$,
with amplitude $-1$,
\item
split-join, $\ldots 01010 \ldots \leftrightarrow \ldots 01100 \ldots$,
$\ldots 01010 \ldots \leftrightarrow \ldots 00110 \ldots$, with amplitude 
$x$,
\item
partner swap, $\ldots 011010 \ldots \leftrightarrow \ldots 010110 \ldots$,
with amplitude $1$,
\item
potential, given by $x^2\widetilde{N}_s + \widetilde{N}_p$,
where $\widetilde{N}_s$ counts the number of sites where, within  
${\cal H}_2$, a single fermion with empty neighboring sites can 
be put in or taken out, and, similarly, $\widetilde{N}_p$ counts the 
number of sites where a fermion can be put in or taken out, such as 
to convert a pair to a single fermion or vice versa. 
\end{itemize}

As an example of the potential, 
the diagonal term for the configuration $0011010$ on a 7-site 
open chain will be $2x^2+3$.
The potential can be rewritten 
in terms of the number of singlets $N_s$, the number of pairs $N_p$,
and the number of ``nearest'' neighbor states, where $N_{ss}$, $N_{sp}$
and $N_{pp}$ counts the number of times singlets or pairs are separated
by only one site in between. We have for $N$ sites
\begin{equation}
 x^2\widetilde{N}_s + \widetilde{N}_p = x^2N - (2x^2-2)N_s - (4x^2-2)N_p
+(x^2-2)N_{ss} + (x^2-1)N_{sp} + x^2N_{pp}\ .
\label{M2potential}
\end{equation}
Clearly, $x=0$, $x=1$, $x=\sqrt{2}$, 
and $x\to \infty$ are special in the sense that
some of the amplitudes drop out and/or become equal to each other.

The general discussion of section \ref{sec:WittenIndex} 
applies to this case,
so we already know the Witten index for any $M_2[x]$: $W_2=3$
for a chain of length $N=4n$, so we know
that there are at least three ground states.
Based on numerical analysis, we expect 
that for general $x\neq 0$, there are precisely three $E=0$ ground 
states at fermion number $f=N/2$.

For $x=0$, the situation is quite different and there are many 
more $E=0$ ground states,  although for $N=4n$ there must be three more
bosonic ground states than fermionic ones, so that
the Witten index remains three. 
A special feature at $x=0$ is that the split-join amplitude
in the Hamiltonian vanishes. This means that
both $N_s$
and $N_p$  are conserved individually,
with the total fermion number equal to $f=N_s+2N_p$.
We observe the following
ground states in $M_2[x=0]$:
\begin{itemize}
\item
the empty state with $N_p=N_s=0$, $t=1$ 
\item
at $t=1$: the unweighted sum of all states with $N_p=1$, $N_s$ even
\item
at $t=1$: the unweighted sum of all states with $N_p=0$, $N_s$ odd 
\item
for $N=4n$, at $t=\pm {\rm i}$, the states $|1010\ldots 10\rangle
 \pm {\rm i} \ |0101\ldots 01\rangle$ 
\item
for $N=4n+2$, at $t=-1$, the state $|1010\ldots 10 \rangle- \ |0101\ldots 01
\rangle$
\end{itemize}
Note that the supersymmetry gives a nice way of finding the ground states:
it is often easier to find states annihilated by
both $Q^+$ and $Q^-$ directly, instead of those with zero eigenvalue of $H$.

An unweighted sum over states amounts to having bare particles of
momentum zero.  Thus the ground states at $x=0$ include fermion
condensates with any fermion number: the chemical potential and
interactions compensate for the fermi repulsion.  In the model $M_1$,
the fermi level is $|p_f|=\pi/3$.  That model has the special property
that multiple fermions are allowed with the same bare momentum equal
to the fermi level. The same property holds in $M_2[x=0]$, but even
more remarkably here, the fermi level here corresponds to $p_f=0$, a
state with zero bare energy. This is why one can have an arbitrary
number of such particles in the ground state, as long as the periodic
boundary conditions are satisfied.  At any finite temperature,
however, there are no $p=0$ particles in the state minimizing the free
energy: adding a $p=0$ particle to a given state reduces the entropy
without decreasing the energy, and so raises the free energy.

\subsection{Relation with the $su(2|1)$ symmetric $tJ$ model when $x=0$}

We earlier discussed a mapping between the simplest supersymmetric 
lattice model $M_1$ and the XXZ chain at $\Delta=-1/2$. We can find
a similar mapping for $M_2[x=0]$. Rather surprisingly, this turns out to lead
to a well-known integrable model, the so-called $tJ$ model at its
integrable ferromagnetic point $J=-2$, $t=1$.

First, it is useful to give the operators $Q^\pm$ and $H$ for $x=0$ in 
terms of the creation and annihilation operators for single fermions 
($s^{\dagger}$ and $s$) and for the pairs ($p^{\dagger}$ and $p$). 
Expressed in terms of the basic fermion annihilation operators we have 
$s_i = d_i = {\cal P}_{i-1}c_i{\cal P}_{i+1}$ at lattice site $i$.
The pairs are located at the sites of the dual lattice, of which the
indices take half-integer values: 
$p_{i+1/2}  = {\cal P}_{i-1}c_{i} c_{i+1}{\cal P}_{i+2}$.
Singlets must be 2 sites from each other, 
pairs must be 3 sites from each other, and a pair and a singlet must 
be 5/2 sites from each other. (One can thus think of this as a system 
of hard-core dimers and trimers.) The supercharges in this basis are 
quite simple: they just change a singlet to a pair, and vice versa, 
namely
$$Q^+=\sum_{i=1}^N (p_{i-1/2}^\dagger - p_{i+1/2}^\dagger)s_i
 \qquad\qquad Q^-=\sum_{i=1}^N (s_{i}^\dagger - s_{i+1}^\dagger)
p_{i+1/2}\ .$$ 
The Hamiltonian is
\begin{eqnarray}
\nonumber
H_2[x=0] &=& 2N_s + 2N_p + \sum_{i=1}^N \big[
(-s^\dagger_{i+1} s_i^{}\ -\ p^\dagger_{i+1/2} p_{i-1/2}^{}
\ +\ p^\dagger_{i+1/2} p_{i+5/2}^{} s^\dagger_{i+3} s_i^{}\ +\ h.c.)\\
&& \qquad\qquad 
\ -\  2 s^\dagger_{i+2} s_{i+2}^{} s^\dagger_{i} s_{i}^{}
\ -\  p^\dagger_{i+5/2} p_{i+5/2}^{} s^\dagger_{i} s_{i}^{}
\ -\  p^\dagger_{i-5/2} p_{i-5/2}^{} s^\dagger_{i} s_{i}^{} \big].
\label{Hpair}
\end{eqnarray}
In these formulas, the action of $Q^\pm$ and $H$ is restricted to the
$k=2$ Hilbert space ${\cal H}_2$. The various amplitude listed in
above are in agreement with (\ref{Hpair}).  This model is
reminiscent of the $tJ$ model in that it involves two species of
particles, with four-particle interactions. 
Note also that in this Hamiltonian
the chemical potentials are
negative and the interactions are attractive,  
both in contrast to the Hamiltonian for $M_1$.
It thus resembles a ferromagnetic $tJ$
model; we give the precise mapping next.

The $tJ$ model describes fermions with spin hopping
on a lattice. No double occupancy is allowed, and the interaction between
nearest-neighbor fermions is of Heisenberg type. Thus we have
three possible states at each site, ($\uparrow$, $\downarrow$, hole).
Comparing with section (\ref{sec:m1xxz}), we again map an edge between 
empty sites to an up-spin, and a single occupied site to a down-spin. 
In addition, we map a pair to an empty state (hole) in the corresponding 
spin model. In the resulting state of the spin model we include a minus 
sign for every pair (hole at $j$, down-spin at $k$) with $j<k$. In terms 
of $N$, $N_p$ and $N_s$, the parameters of the spin model are
\be
N_\uparrow=N-2N_s-3N_p, \quad N_\downarrow=N_s, \quad N_h=N_p
\ee
and the spin model lives on a chain with $L=N-N_s-2N_p$ sites.
Converting the Hamiltonian (\ref{Hpair}) to act on the spin model produces
precisely the Hamiltonian of the $tJ$ model, with $J=-2$ and $t=1$:
\be
H_2[x=0] \leftrightarrow 
H_{tJ}[t=1,J=-2] + 2 N_h \ ,
\ee
where the $tJ$ Hamiltonian is given by (we follow the conventions of 
\cite{EK})
\begin{eqnarray}
H_{tJ} &=&
-t \sum_{j=1}^L \sum_{\sigma=\uparrow,\downarrow}
\left[ Q^\dagger_{j,\sigma} Q_{j+1,\sigma}
       + Q^\dagger_{j+1,\sigma} Q_{j,\sigma} \right]
\nonumber\\ &&
+ J \sum_{j=1}^L 
\left[ S^{\rm z}_j S^{\rm z}_{j+1}
       + \frac{1}{ 2} ( S^+_j S^-_{j+1} + S^-_j S^+_{j+1} )
       - \frac{1}{ 4} n_j n_{j+1} \right] \ .
\end{eqnarray}
The operators $Q^\dagger_{j,\sigma}$,
$Q_{j,\sigma}$ and $S^a_j$ and $n_j$ are bilinears
of the $tJ$ fermion annihilation/creation operators 
$c_{\sigma,j},c_{\sigma,j}^\dagger$ 
and the bosonic hole annihilation/creation operators 
$b_j,b^\dagger_j$ at site $j$:
\begin{eqnarray*}
&& 
Q^\dagger_{j,\sigma} = c^\dagger_{j,\sigma} b_j (1 - n_{j,-\sigma})
\ , \quad
Q_{j,\sigma} = b^\dagger_j c_{j,\sigma} (1 - n_{j,-\sigma})
\\ &&
S^+_j=c^\dagger_{\uparrow,j}c^{}_{\downarrow,j}
\ , \quad
S^-_j=c^\dagger_{\downarrow,j}c^{}_{\uparrow,j}
\ , \quad
S^{\rm z}_j = \frac{1}{2}(n_{j,\uparrow}-n_{j,\downarrow})
\\ &&
n_{j,\sigma} = c^\dagger_{\sigma,j} c_{\sigma,j}
\ , \quad
n_j= n_{j,\uparrow}+ n_{j,\downarrow} \ .
\end{eqnarray*}
These operators all act within the physical space defined by
$n_{j,\uparrow}+n_{j,\downarrow}+b_j^{\dagger} b_j=1$ for all sites $j$.

As for the mapping $M_1 \leftrightarrow$ XXZ, the resulting $tJ$
model is subject to twisted boundary conditions, with the twist
depending on the eigenvalue $t$ of the translation operator. 
In particular
\be 
c^\dagger_{L+1,\uparrow} = c^\dagger_{1,\uparrow} 
\qquad
c^\dagger_{L+1,\downarrow} = 
  (-1)^{1+N_\downarrow+N_h} t^{-1} c^\dagger_{1,\downarrow}
\qquad
b^\dagger_{L+1} = (-1)^{N_\downarrow} t^{-2} b^\dagger_{1} \ .
\label{twistedBC}
\ee

The $tJ$ model at $J=\pm 2t$, with periodic boundary conditions, 
is invariant under the graded Lie algebra symmetry $su(2|1)$. 
On a given site, the $su(2|1)$ 
symmetry rotates the three possible states on the site
amongst each other. The algebra is generated by $Q_\sigma,\
Q_\sigma^\dagger,\ S^i$ and $n_\sigma$.
Since an electron is fermionic
and the empty site bosonic, some of these generators
are fermionic, and some are bosonic. For
this reason graded Lie algebra symmetries are usually called
`supersymmetries' in the condensed-matter literature.
This terminology is unfortunate, 
as the $su(2|1)$ symmetry of the `supersymmetric' $tJ$ model is unrelated 
to the (spacetime) supersymmetry described in this paper. Because of the
twist involved in the mapping
$M_2[x=0] \leftrightarrow tJ$,
there is no Hamiltonian in finite volume which
has both the supersymmetry and the graded Lie algebra
symmetry $su(2|1)$: in the fermion model $M_2[x=0]$ the $su(2|1)$ structure
is hidden, and in the $tJ$ model there is a hidden supersymmetry. 

The mapping does elucidate some aspects of $M_2[x=0]$. In the
absence of twists in the boundary conditions, 
the $tJ$ model with ferromagnetic sign 
of the spin-spin interaction has an $E=0$ ground state multiplet,
which is a multiplet of dimension $2L+1$ of the $su(2|1)$ symmetry,
at $t=1$. It consists of two $su(2)$ multiplets, one of dimension 
$L+1$ with highest weight state with $N_\uparrow=L$, $N_\downarrow=N_h=0$, 
the other of dimension $L$ and highest weight state with 
$N_\uparrow=L-1$, $N_\downarrow=0$, $N_h=1$. 
In the presence of the twisted boundary conditions, eq. (\ref{twistedBC}),
the states with $N_\downarrow=0$ or $N_h+N_\downarrow$ odd survive as 
$E=0$ ground states.
Mapping back to $M_2[x=0]$ produces the $E=0$ ground states with
$t=1$ and $N_p=N_s=0$, ($N_p=0$, $N_s$ odd) and ($N_p=1$, $N_s$ even)
that we discussed in section \ref{sec:HamGS}.

\subsection{Relation with integrable spin-1 XXZ model when $x=\sqrt{2}$}
\label{sec:spin1XXZ}

The Hamiltonian for $M_2[x]$ for $x\ne 0$
does not conserve $N_s$ and $N_p$ individually, but only the fermion
number $N_s + 2N_p$. As a result, it is quite complicated. 
Nevertheless, at $x=\sqrt{2}$, it is integrable. We show this is so
in this section by mapping it onto an integrable spin-1 XXZ chain.

This mapping proceeds similarly to the above two mappings. Because the
singlets and pairs in $M_2[x]$ take up multiple sites, the
correspondence is between $M_2[x]$ on the periodic chain with $N$
sites and the spin-1 XXZ chain with $L=N-f$ sites and twisted boundary
conditions. The states in the spin-1 XXZ chain are denoted $+$, $0$
and $-$, with the Hamiltonian conserving $N_+ - N_-= L- N_0 - 2
N_-$.

We identify the empty state in $M_2[x]$ with the
state with all spins $+$ in the spin-1 chain. The singlets are
identified with spin $0$ and the pairs are identified with spin $-$ in
the same fashion as in the $tJ$ model above.
The four off-diagonal terms in the Hamiltonian for $M_2[x]$ 
listed at the beginning
of this section then correspond respectively to the processes
\begin{eqnarray*}\, . \, . \, . +\,0\, . \, . \, .
 &\leftrightarrow&\, . \, . \, .\ 0\, + \, . \, . \, . \\
\, . \, . \, . +-\, . \, . \, . &\leftrightarrow & \, . \, . \, . -+ \, . \, . \, . \\
\, . \, . \, . \ \, 0\ \, 0\, . \, . \, . &\leftrightarrow &\, . \, . \, . +- \, . \, . \, . \qquad\hbox{and}\qquad
\, . \, . \, .\ 0\ 0 \ .\, .\, . \leftrightarrow \, .\, .\, . -+ .\, .\, . \\
\, . \, . \, . -\, 0\, . \, . \, . &\leftrightarrow &\, . \, . \, .\ 0\, - \, . \, . \, . 
\end{eqnarray*}
The coefficient of these terms in the $M_2[x]$ Hamiltonian
are respectively $x^2-1,-1,x$ and $1$. 
Thus if we demand that the spin-chain Hamiltonian
satisfy a $\bf{Z}_2$ symmetry under $+\leftrightarrow -$, this fixes
the coupling in the corresponding model $M_2[x]$ to be $x=\sqrt{2}$.
At this value of $x$ the diagonal part of the Hamiltonian, denoted by
$ 2\widetilde{N}_s + \widetilde{N}_p $ in  $M_2[x]$, reduces to the
simple expression $$2 L - N_{+-} - N_{-+}$$ in the corresponding spin chain,
also manifestly  $+\leftrightarrow -$ symmetric.
This means that $M_2[\sqrt{2}]$ indeed corresponds to a spin-1 XXZ
chain.

To show that $M_2[\sqrt{2}]$ is integrable, we now show that the corresponding
spin-1 XXZ chain is integrable.
Since the $SU(2)$ symmetry of the spin-1 chain is broken to $U(1)\times 
{\bf Z}_2$, there are many possible nearest-neighbor couplings. The model is
integrable only in a one-parameter family of these couplings \cite{FZ}.
These couplings can be obtained from the spin-1/2 XXZ chain by implementing
a procedure called fusion on the related six-vertex model with
$\Delta=-\cos(\eta)$ \cite{KRS}. One
finds that the integrable spin-1 Hamiltonian on $L$ sites 
can be written as
\begin{eqnarray*}
H_{{\rm XXZ}-1}&=&\sum_{j=1}^{L} \Big[
 \tilde{S}^+_j \tilde{S}^-_{j+1}\,  + \, \tilde{S}^-_j \tilde{S}^+_{j+1}
\, - \, (\tilde{S}^+_j)^2 (\tilde{S}^-_{j+1})^2 
\, - \, (\tilde{S}^-_j)^2 (\tilde{S}^+_{j+1})^2
\, - \, \tilde{S}^+_j \tilde{S}^-_j \tilde{S}^-_{j+1} \tilde{S}^+_{j+1}\\
&&
\, - \, \tilde{S}^-_j \tilde{S}^+_j \tilde{S}^+_{j+1} \tilde{S}^-_{j+1}
\, +\, (1-2\cos(\eta))\{S^{\rm z}_j S^{\rm z}_{j+1},
\tilde{S}^+_j \tilde{S}^-_{j+1} + \tilde{S}^-_j \tilde{S}^+_{j+1}\} \\
&&+\cos(2\eta) S^{\rm z}_j S^{\rm z}_{j+1} 
\,-\, \cos(2\eta)(S^{\rm z}_j)^2 (S^{\rm z}_{j+1})^2 \, + \, 2(\cos(2\eta)-1)(S^{\rm z}_j)^2
\Big]
\end{eqnarray*}
where
$$S^{\rm z} = \begin{pmatrix}
1&0&0\cr 0&0&0\cr 0&0&-1
\end{pmatrix}\qquad
\tilde{S}^+ = \begin{pmatrix}
0&1&0\cr 0&0&1\cr 0&0&0
\end{pmatrix}\qquad
\tilde{S}^- = \begin{pmatrix}
0&0&0\cr 1&0&0\cr 0&1&0
\end{pmatrix}\ .$$
The matrices $\tilde{S}^\pm$
are defined without the $\sqrt{2}$ appearing in the usual $SU(2)$
spin-1 generators. When $\eta=0$, $SU(2)$ invariance reappears; when
$\eta=\pi/4$, $\cos(2\eta)$ vanishes and the Hamiltonian simplifies.

It is now straightforward to verify that when $\eta=\pi/4$, the above
mapping takes the Hamiltonian for $M_2[\sqrt{2}]$ to $H_{{\rm XXZ}-1}
+ 4 L$. Note again that the shift $4L$ is a constant in the XXZ model,
but not in the original fermion chain.  To find an equivalent model
where the shift is independent of $L$, one can include a magnetic
field term $-\sum_{j=1}^L{\cal H}S^z_j$ in the spin-chain Hamiltonian. Then the
Hamiltonian for $M_2[\sqrt{2}]$ maps on to $H_{{\rm XXZ}-1} +
2N$ with ${\cal H}=-2$.

The Hamiltonians of $M_2[\sqrt{2}]$ and the spin-1 XXZ chain at
$\Delta=-1/\sqrt{2}$ are therefore identical, up to a shift and a
twist in the latter. As with the $tJ$ model, when one takes into
account the periodic boundary conditions, the spin-1 model must have
twisted boundary conditions. The diagonal terms $ - N_{+-} - N_{-+}$ clearly
show that the spin model is antiferromagnetic.  Despite the fact that
this Hamiltonian is relatively complicated, implementing the Bethe
ansatz is relatively straightforward.  The fusion procedure ensures
that the Bethe equations will be the same as for the spin-1/2 XXZ
model with $\Delta =-1/\sqrt{2}$; only the expression for the energy
will change. The ground states for periodic boundary conditions will
have $E_0=0$, so this means it is likely that ${\cal Q}_0$ for
$M_2[\sqrt{2}]$ can be calculated explicitly as for $M_1$.

Just as the spin-1/2 XXZ model at $\Delta=-1/2$ is described by
an ${\cal N}=(2,2)$ supersymmetric conformal field theory in the
continuum limit, so is the integrable spin-1 XXZ model at 
$\Delta=-1/\sqrt{2}$. That this is the value of $\Delta$ which
gives a supersymmetric field theory can be checked by studying the
system at finite temperature. It is straightforward to derive
the thermodynamic Bethe ansatz (TBA) equations giving the
free energy in the continuum limit of this theory (see e.g.\ \cite{TBA}). 
They are identical to those for the
${\cal N}=(2,2)$ superconformal field theory with $c=3/2$, as can
be shown by taking the massless limit of the TBA equations for the
supersymmetric field theory with Landau-Ginzburg potential $X^4-X^2$
found in \cite{FI}.

\subsection{Nested BA solution at $x=0$}

We can solve the models $M_2[x=0]$ and $M_2[x=\sqrt{2}]$ 
by using the Bethe ansatz. Although
this solution can be inferred from the mappings to the $tJ$ model 
and the spin-1 XXZ model with
twisted boundary conditions, it is instructive to solve the model
directly. We do so in this section for $M_2[x=0]$.

Since there are two different kinds of particles, the Bethe ansatz
is more complicated. Particles of different species can
hop over each other, so we need to also track of the ordering of a given
state. We denote the locations of the
singlets by $i_1,i_2,\dots i_{N_s}$,
and the pairs by $i_{N_s+1},i_{N_s+2},\dots i_{M}$,
where $M=N_s+N_p$ is the total number of particles.
A particular ordering $i_{Q1}<i_{Q2}<\dots <i_{QM}$, 
is labeled by $Q$, a permutation of $1\dots M$. The ansatz
for the wavefunction is then
$$\varphi^{(N_s,N_p)}(i_1\dots i_M) = \sum_{P,Q} A_{P,Q}\,
\mu_{P1}^{i_{Q1}} \mu_{P2}^{i_{Q2}}\dots \mu_{PM}^{i_{QM}}
$$
where $P$ is another permutation of $1\dots M$. 
If all the particles are of one type, then we don't need
the sum over $Q$: the symmetry or antisymmetry of the wavefunction
relates the wavefunctions for different orderings of the particles.
The eigenvalue
of the translation operator is
$$t= \prod_{j=1}^M (\mu_j)^{-1}$$
as before. 
If this is an eigenstate of $H$, the eigenvalue is
$$E=2M - \sum_{j=1}^M \left[\mu_j + (\mu_j)^{-1}\right].$$

Let us first look at the system with only pairs present. 
We see from the Hamiltonian (\ref{Hpair}) that this is a free
system, except for the hard-core repulsion. The repulsion means
that for any pair of particles, an eigenstate must satisfy
$$\varphi^{(0,N_p)}(\dots,i+1,i+3,\dots)+
\varphi^{(0,N_p)}(\dots,i,i+2,\dots)=0$$
for any $i$.
This means that
$$\sum_P A_P\left(\mu_{Pj}^{i+1}\mu_{P(j+1)}^{i+3} +
\mu_{Pj}^{i}\mu_{P(j+1)}^{i+2}\right)=0$$  
for all $i$ and $j$.
This is satisfied if
$$\frac{A_P'}{A_P}= -
\frac{\mu_{P(j+1)}^2 + \mu_{Pj}\mu_{P(j+1)}^3}
{\mu_{Pj}^2 + \mu_{P(j+1)}\mu_{Pj}^3}  = 
-\frac{\mu_{P(j+1)}^2}
{\mu_{Pj}^2} $$
This yields the Bethe equations
$$\mu_j^{N-2M}t^{-2} = (-1)^{M-1}$$
Thus as before, the hard-core repulsion effectively reduces the size
of the system and inserts a twist.

The system with only singlets is very similar to the system $M_1$.
Their Hilbert spaces are identical, while the Hamiltonian for $M_2[x=0]$
has twice the potential, and is multiplied by an overall minus sign.
The singlet-only subsector thus corresponds to a
hard-core fermion representation of the XXZ model at $\Delta=1$,
the ferromagnetic Heisenberg model. This does not mean that the 
ferromagnetic Heisenberg model is supersymmetric, because the supersymmetry
converts singlets to pairs, and takes the theory out of the singlet-only
sector. Repeating the earlier Bethe ansatz analysis with these
slight modifications, one has
$$\varphi^{(N_s,0)}(\dots,i+1,i+2,\dots)+\varphi^{(N_s,0)}(\dots,i,i+1,\dots)-
2\varphi^{(N_s,0)}(\dots,i,i+2,\dots)=0$$
These vanish if
$$\frac{D_{P'}}{D_{P}}=-\frac{y}{x}
\frac{xy +1 - 2y} {xy +1 - 2x}.$$
where henceforth we denote $\mu_{Pj}=x$ and $\mu_{P(j+1)}=y$
for simplicity. We have denoted 
the coefficient $D_P\equiv A_{P,(12\dots M)}$ here.
The Bethe equations are
$$
\mu_k^{N-M} t^{-1} = \prod_{j\ne k} \frac{\mu_k\mu_j +1 -2\mu_k}
{\mu_k\mu_j +1 -2\mu_j}.$$
The correspondence with the Heisenberg model is not surprising in light
of the map to the $tJ$ model; in the latter with an electron on
every site, the hopping term is zero, and the model reduces to the
Heisenberg model, with the spin up and down electrons playing the
role of the Heisenberg spins. In the map of the $M_2[x=0]$ to the 
$tJ$ model, a state with only singlets in the former indeed
maps to a state with an electron on every site in the latter.

These two special cases in fact yield all the two-particle excited states,
since we can use $Q^+$ on $\Phi^{(2,0)}$ and $Q^-$ on the $\Phi^{(0,2)}$
to obtain any excited state with $N_s=N_p=1$.
It is still useful to apply the Bethe ansatz here, to derive
the $R$-matrix necessary for the nesting. 
Let $i_1$ be the location of the singlet, and
$i_2$ be the location of the pair. We then write the Bethe ansatz as
$$ \phi^{(1,1)}(i_1,i_2)=
\begin{cases}
B_{12} x^{i_1}y^{i_2} + B_{21} y^{i_1}x^{i_2}& i_1 <i_2\\
C_{12} x^{i_2}y^{i_1} + C_{21} y^{i_2}x^{i_1}& i_1 >i_2
\end{cases}
$$
In the previous notation, $B_{12}\equiv A_{(12),(12)}$,
$B_{21}\equiv A_{(21),(12)}$, $C_{12}\equiv A_{(12),(21)}$, 
$C_{21}\equiv A_{(21),(21)}$.
This is an eigenstate of the Hamiltonian (\ref{Hpair}) if
\begin{eqnarray*}
0&=&
\varphi^{(1,1)}(i+1,i+5/2)+\varphi^{(1,1)}(i,i+3/2)-
\varphi^{(1,1)}(i,i+5/2) + \varphi^{(1,1)}(i+3,i+1/2)\\
&=&\varphi^{(1,1)}(i-1,i-5/2)+\varphi^{(1,1)}(i,i-3/2)-
\varphi^{(1,1)}(i,i-5/2) + \varphi^{(1,1)}(i-3,i-1/2)
\end{eqnarray*}
These equation can be verified with the explicit $\varphi^{(1,1)}$
found by applying $Q^{-}$ to $\phi^{(0,2)}$ and $Q^+$ to $\phi^{(2,0)}$.
Substituting the explicit form of $\varphi^{(1,1)}$ into the first
of these equations yields
$$0=B_{12}y^{3/2}(1+xy-y)
+ B_{21}x^{3/2}(1+xy-x)
+ C_{12}x^{1/2}y^3  + C_{21}y^{1/2}x^3 
$$
while second one yields
$$0=C_{12}y^{3/2}(1+xy-y)
+ C_{21}x^{3/2}(1+xy-x)
+ B_{12}x^{-1/2}y^2  + B_{21}y^{-1/2}x^2.$$ 
Since there are two linear equations with four unknowns, we can solve for
$B_{21}, C_{21}$ in terms of $B_{12}, C_{12}$, yielding
$$
\begin{pmatrix}
B_{21}\\ C_{21}
\end{pmatrix}
=\frac{1}{xy+1-2x}\frac{y^{3/2}}{x^{3/2}}
\begin{pmatrix}
-(x-1)(y-1)& x^{1/2}y^{1/2} (x-y)\\
 x^{-1/2}y^{-1/2} (x-y) & - (x-1)(y-1)
\end{pmatrix}
\begin{pmatrix}
B_{12}\\ C_{12}
\end{pmatrix}
$$

Putting all this together gives us the $R$-matrix of the theory.
Denoting the vector
$\xi_{P} = (A_{P},B_{P},C_{P},D_{P})$, the $R$-matrix is defined
as $\xi_{21}=R(x,y) \xi_{12}$. Here
the matrix $R$ is
$$R=-\frac{1}{xy+1-2x}\frac{y^{3/2}}{x^{3/2}}
\begin{pmatrix}
\frac{y^{1/2}}{x^{1/2}}(xy+1-2x)&0&0&0\\
0&(x-1)(y-1)& -(xy)^{1/2} (x-y)&0\\
0& -(xy)^{-1/2}(x-y) & (x-1)(y-1)&0\\
0&0&0&\frac{x^{1/2}}{y^{1/2}}(xy+1-2y)
\end{pmatrix}
$$
This will look a little more familiar if we define
$$x\equiv\frac{\theta_1+{\rm i}\pi/2}{\theta_1-{\rm i}\pi/2}\qquad\qquad
y \equiv\frac{\theta_2+{\rm i}\pi/2}{\theta_2-{\rm i}\pi/2}$$
so
$$R=-\frac{1}{\theta+{\rm i}\pi}\frac{y^{3/2}}{x^{3/2}}
\begin{pmatrix}
\frac{y^{1/2}}{x^{1/2}}(\theta+{\rm i}\pi)&0&0&0\\
0&{\rm i}\pi& (xy)^{1/2}\theta &0\\
0& (xy)^{-1/2}\theta & {\rm i}\pi&0\\
0&0&0&\frac{x^{1/2}}{y^{1/2}}({\rm i}\pi-\theta)
\end{pmatrix}
$$
where $\theta\equiv\theta_1-\theta_2$.
This is the $R$-matrix which gives the Boltzmann weights for
a particular six-vertex model in an electric field; the extra factors
of $x$ and $y$ in the last equation cancel out of the resulting partition
function. The $R$-matrix 
satisfies the inversion relation $R(x,y)R(y,x)=1$ as it should. 

This $R$-matrix is characteristic of those arising in ${\cal N}$=2
supersymmetric field theories, in that it satisfies what
is known as the free-fermion condition.
Denoting the six non-zero elements of the $R$-matrix in order as
$a,c,b,\widetilde{b},\widetilde{c},\widetilde{a}$, it
is easy to check that they satisfy
$a\widetilde{a}+b\widetilde{b}-c\widetilde{c}=0$.  
In supersymmetric field theories, the supersymmetry constrains
the elements of the two-particle $S$ matrix to obey this relation
\cite{FI}. We have found other integrable ${\cal N}$=2 
supersymmetric
lattice models, and their $R$-matrices also satisfy the free-fermion
condition. Presumably, the supersymmetry constrains the $R$-matrix
of our lattice models to be of this form,
but we do not have a general proof of this observation.
(The name free-fermion condition is a bit confusing: it means
that a classical lattice model with 
Boltzmann weights given by this $R$-matrix
can be solved by using Pfaffians,
without recourse to the Bethe ansatz. Such a lattice model 
is equivalent to a free-fermion theory, but our quantum theories are not:
the $R$-matrix appears in the nested Bethe ansatz, not as a Boltzmann weight.)

This $R$ matrix satisfies the Yang-Baxter equation, 
so we can apply
the nested Bethe ansatz to find the energy eigenstates for general
$(N_s,N_p)$. Since this procedure is standard, we omit the details
here. An excellent reference is the original paper by Yang \cite{Yang}.
Associated with each particle is a variable $\mu_i$,
$i=1\dots N_s+N_p$, so that $\ln(\mu_i)/{\rm i}$ is its bare momentum.
We then need the ``magnon'' variables $\nu_k$ to account for
the extra degrees of freedom of allowing a given particle to be
either a singlet or a pair. If we associate the $\nu_k$ with
singlets in a background of pairs, we then have $k=1\dots N_s$. 
The resulting nested
Bethe equations are
\begin{eqnarray}
\mu_i^N &=& - \prod_{k=1}^{N_s} \frac{ (\mu_i - \nu_k)}{\mu_i(\mu_i \nu_k + 1 - 2 \nu_k)} 
             \prod_{j=1}^{N_s+N_p} \frac{\mu_i^2}{\mu_j^2}
\nonumber\\
  1 &=&  \prod_{i=1}^{N_s+N_p} \frac{ (\mu_i - \nu_k)}{\mu_i(\mu_i \nu_k + 1 - 2 \nu_k)} 
\ .
\end{eqnarray}
Each solution of these equations represents an eigenstate of the Hamiltonian
with the energy
\be 
E = \sum_{i=1}^{N_s+N_p} \left[2 - \mu_i - \mu_i^{-1} \right].
\label{BAenergy}
\ee
If we instead associate the magnons with the pairs so that $k=1\dots N_p$,
the Bethe equations 
\be \mu_i^N = \prod_{j=1}^{N_p+N_s} 
\frac{\mu_i (\mu_i \mu_j + 1 - 2 \mu_i)}{\mu_j (\mu_i \mu_j + 1 - 2 \mu_j)}
\prod_{k=1}^{N_p} \frac{\mu_i(1 - \mu_i \nu_k)}{\mu_i + \nu_k - 2 \mu_i \nu_k}
\label{ba2.1}
\ee
\be 1= \prod_{j=1}^{N_p+N_s}\frac{\mu_j(1 - \mu_j \nu_k)}{\mu_j + \nu_k - 2 \mu_j \nu_k}
\ee
with $E$ again given by (\ref{BAenergy}).
In either set of Bethe equations, the supersymmetry doublets
are states with identical $\mu_j$ and $\nu_k$, save for one extra
magnon with $\nu_k=0$.

These two sets of Bethe equations have both arisen in the $tJ$ model:
our second set corresponds to the Sutherland or `BFF' set\cite{Su,EK},
while our first solution is the `FBF' set first given in
\cite{EK}. The two sets were shown to be equivalent in \cite{EK}.
Comparing the $tJ$ equations to those for $M_2[x=0]$
confirms the twist that we already described, and the relation
$L=N-N_s-2N_p$. We stress that, while in the end there is agreement
between the Bethe equations, the underlying algebraic structures are
very different: our analysis here is anchored on $4 \times 4$
supersymmetric $R$-matrices, while the solutions of the $tJ$ model
involve $9 \times 9$ $R$-matrices associated to the graded Lie algebra
$u(2|1)$.

\section{General $M_k$ models and the CFT connection}

We have already mentioned in
sections \ref{sec:holonsft} and \ref{sec:spin1XXZ} 
that in the continuum limit, 
the models $M_1$ and $M_2[\sqrt{2}]$ are each
described by ${\cal N}=(2,2)$ supersymmetric conformal field
theories. The operator content of the former case is discussed in detail in
ref.\ \cite{FSd}; the analysis of the latter case is very similar.
The simplest superconformal field theories (SCFTs) are called minimal
models. They are indexed by an integer $k$, and have a central charge
$3k/(k+2)$. The field theories describing the continuum limit
of the models $M_1$ and $M_2[\sqrt{2}]$ are the first two superconformal
minimal models.

This correspondence makes it highly plausible that the models
$M_k[\{x_i\}]$ are also related to minimal models.  We thus conjecture
that for a specific choice of the parameters $x_1,\ldots,x_{k-1}$, the
model $M_k$ has a continuum limit described by $k$-th ${\cal N}=(2,2)$
superconformal minimal model.  One piece of evidence for this
conjecture is that the value of the Witten index for $M_k$ on a closed
chain with $N=n(k+2)$ is $|W_k|=k+1$, in agreement with the Witten
index in the $k$-th minimal model \cite{LW}.

Another, very different, piece of evidence stems from the fact that
there exist a finitization procedure which relates the Hilbert spaces
of the SCFT and the lattice model. By starting from the chiral Hilbert
space of the $k$-th ${\cal N}=(2,2)$ SCFT, one produces finite spaces,
graded by fermion number $f$, with dimensionalities identical to those
of ${\cal H}_k^{(f)}$. The finitization scheme employs a
quasi-particle basis of the ${\cal N}=(2,2)$ CFT, which is related to the
so-called spinon basis of the $k$-th minimal model of $su(2)$
invariant CFT \cite{BLS}.

We also note that the structure of the Hilbert spaces ${\cal H}_k$,
involving clustering of up to $k$ fundamental fermions, seems natural
from the point of view of the $Z_k$ parafermions underlying the $k$-th
minimal ${\cal N}=(2,2)$ SCFT. For example, in \cite{RR}, these CFTs have
been related to quantum Hall states with a order-$k$ clustering
property similar to ours.

The relation between the models $M_1$ and $M_k$ is in a way similar
to that between the spin $S=1/2$ and spin $S=k/2$ nearest neighbor 
Heisenberg models on a 1D chain. In the latter case, there are $k-1$
adjustable parameters in the Hamiltonian, which can be tuned to 
produce a critical model having the $k$-th minimal $su(2)$ invariant
CFT for its continuum limit. The latter CFTs are closely related
to the $k$-th minimal ${\cal N}=(2,2)$ SCFTs.

Finally, the presence of supersymmetry enables computations like that
of sect.\ \ref{sec:susyGS} even at finite temperature. In the simplest
case, at a special value of (imaginary) chemical potential, the
$Q$-function for $M_1$ in the continuum limit turns into a Bessel
function \cite{PF}. This simplification has useful consequences,
because the $Q$-functions for a variety of integrable field theories
have been shown to be useful in deriving the spectral determinant for
the Schr\"odinger equation with certain potentials. This
simplification has been shown to take place in many cases, including
the integrable spin-$k/2$ spin chains at their supersymmetric points
$\Delta=-\cos(\pi/(k+2))$ \cite{Patrick}.  (The fact that these points
are described by supersymmetric field theories can be shown by
extending the arguments at the end of sect.\ \ref{sec:spin1XXZ}). One
might wonder if the simplification in general is related to the
presence of supersymmetry \cite{Patrick}; indeed, the analysis of
\cite{PF} requires the supersymmetry. We have shown in this paper that
in the spin-$1/2$ and spin-$1$ cases, the supersymmetry is present
even on the lattice. Therefore, the fact that the 
analysis of \cite{Patrick} succeeds for any $k$ is yet another
indication that the models $M_k$ contain a point at which they are
solvable and described by the $k$-th superconformal minimal model.

\vskip 1cm

\noindent
{\bf Acknowledgments}
We thank Oleg Tchernyshyov for informing us of the references \cite{xxzexp}.
This work was supported in part by the foundations FOM and
NWO of the Netherlands, by NSF grant DMR-0104799, and by a 
DOE OJI award.

\end{document}